\documentclass[a4paper,12pt]{article}
\pdfoutput=1 

\usepackage{a4wide}
\usepackage{cite}

\usepackage[T1]{fontenc} 
\usepackage[latin1]{inputenc}

\usepackage{empheq}
\usepackage{amssymb}
\usepackage{amsmath}
\usepackage{amsfonts}
\usepackage{xcolor}

\DeclareMathAlphabet{\mathpzc}{OT1}{pzc}{m}{it}

\numberwithin{equation}{section}

\usepackage[colorlinks]{hyperref}

\hypersetup{
 citecolor=blue,
 linkcolor=blue,
 urlcolor=blue}

\newcommand{\la}{\left\langle}
\newcommand{\ra}{\right\rangle}
\newcommand{\stij}[1]{\tilde s_{#1}}
\newcommand{\lp}{\left(}
\newcommand{\rp}{\right)}
\newcommand{\nn}{\nonumber}

\def \qas {[\alpha_s]}
\def \w {\omega}
\def \KTL {\widetilde{\mathcal K}}
\newcommand{\ep}{\epsilon}
\newcommand{\be}{\begin{equation}}
\newcommand{\ee}{\end{equation}}
\newcommand{\bes}{\begin{equation}\begin{split}}

\def \as {\alpha_s}
\def \asontwopimu {\frac{\alpha_s(\mu)}{2\pi}}
\def \Li {\text{Li}}
\def \gsb {g_{s,b}}
\def \zm {z_{\rm min}}
\def \mq {m_{H}}
\def \wt {\tilde\omega}
\def \G {\mathcal G}

\def \st {\tilde{s}_{12}}

\def \Ca {C_A}
\def \Cf {C_F}
\def \tr {T_R}
\def \nf {n_f}

\def \Em {E_{\rm max}}
\def \d {{\rm d}}

\def \SS {S{\hspace{-5pt}}S}
\def \CC {C{\hspace{-6pt}C}}
\def \FLM {F_{\rm LM}}
\def \FLV {F_{\rm LV}}
\def \FLVV {F_{\rm LVV}}

\def \I {I}
\def \ONLO {\hat{\mathcal O}_{\rm NLO}}

\def \LO {{\rm LO}}
\def \NLO {{\rm NLO}}
\def \NNLO {{\rm NNLO}}

\newcommand{\dg}[1]{[df_{#1}]}
\newcommand{\df}[1]{[df_{#1}]}
\newcommand{\FLMs}[1]{F_{{\rm LM},{#1}}}
\newcommand{\FLVfin}{F_{{\rm LV}}^{\rm fin}}
\newcommand{\FLVVfin}{F_{{\rm LVV}}^{\rm fin}}
\newcommand{\FLVsqfin}{F_{{\rm LV}^2}^{\rm fin}}

\def \qb {{\bar q}}

\begin{document}
\vspace{-5.0cm}
\begin{flushright}
OUTP-19-08P,
CERN-TH-2019-106, TTP-19-023, P3H-19-017
\end{flushright}

\vspace{2.0cm}

\begin{center}
{\large \bf 
  Analytic results for decays of color singlets  to $gg$ and $q \bar q$
  final states at NNLO QCD with  the nested soft-collinear subtraction scheme}\\
\end{center}

\vspace{0.5cm}

\begin{center}
Fabrizio Caola$^{1}$, 
Kirill Melnikov$^{2}$, Raoul R\"ontsch$^{3}$.\\
\vspace{.3cm}
{\it
{}$^1$Rudolf Peierls Centre for Theoretical Physics, Clarendon Laboratory, Parks Road, Oxford OX1 3PU, UK \& 
Wadham College, Oxford OX1 3PN\\
{}$^2$Institute for Theoretical Particle Physics, KIT, Karlsruhe, Germany\\
{}$^3$Theoretical Physics Department, CERN, 1211 Geneva 23, Switzerland\\
}

\vspace{1.3cm}

{\bf \large Abstract}
\end{center}
We present compact analytic formulas that describe the decay of colorless particles to both
$q \bar q$ and $gg$ final states through next-to-next-to-leading
order in perturbative QCD in the context of the nested soft-collinear subtraction scheme.
In addition to their relevance for the description of 
decays like $V \to q \bar q'$, $V= Z,W$, $H \to b \bar b$ and  $H \to gg$,  
these results provide an important building block for calculating
NNLO QCD corrections to arbitrary processes at colliders within the nested soft-collinear subtraction scheme. 

\thispagestyle{empty}

\clearpage
{\hypersetup{linkcolor=black}
\tableofcontents
}
\thispagestyle{empty}
\clearpage

\pagenumbering{arabic}

\allowdisplaybreaks

\section{Introduction}
The development of an efficient and physically transparent subtraction scheme for next-to-next-to-leading order (NNLO) computations in QCD 
is an important problem in theoretical particle physics that attracted a lot of attention recently
\cite{ant,czakonsub,czakonsub4d,Boughezal:2011jf,Cacciari:2015jma,qt1,qt2,njet1,njet2,colorful,
  tackmann:nj:pow,frank:nj:pow,Ebert:2018gsn,magnea,herzog,Caola:2017dug}.
 However, among the many
 subtraction schemes that have been proposed, there is not a single one that is generic, fully local and fully
 analytic (in a sense that 
all the integrated  subtraction terms are available in an analytic form).  Given the 
impressive practical successes of many subtraction schemes in describing physical
processes,
it is unclear whether or not
 locality and analyticity are truly
essential. However,  we believe that it is useful to develop a scheme that is general, physically transparent and
efficient, especially in view of the  need to extend the functionality  of existing subtraction schemes beyond 
$2 \to 2$ processes for forthcoming  LHC applications.  

In Ref.~\cite{Caola:2017dug}, we introduced  the nested soft-collinear subtraction scheme. It
is based on the idea of sector decomposition~\cite{czakonsub} but it relies heavily
on the phenomenon of color coherence in constructing soft and collinear approximations
to matrix elements. This  subtraction scheme is  local by construction; however, initially, 
some  subtraction terms were not known analytically. Recently, this problem was
solved for both the double-soft \cite{max_soft} and triple-collinear \cite{maxtc} subtraction terms so that analytic
results for all double-unresolved subtraction terms are now available. 
Building on that,  in Ref.~\cite{Caola:2019nzf} we  presented analytic results for the production of a color-singlet final
state in hadron collisions obtained within the nested soft-collinear subtraction scheme. In addition to their
phenomenological relevance, we view these results
as building blocks that should, eventually, allow us to describe arbitrary hard processes at hadron colliders through
NNLO QCD. 
Typically, these building blocks are obtained by partitioning the phase space for a particular process
in such a way that only emissions off two hard particles at a time  lead to infra-red and collinear singularities
when integration  over the phase space is attempted.
These hard emittors can be both in the initial or in the final state or
one of them can be in the initial and the other one in the final state. 
When looking  at the problem of constructing a subtraction scheme
from this perspective, the results presented in Ref.~\cite{Caola:2019nzf}  should facilitate the description
of the two initial-state  emittors.

The goal of this paper is to take one further step towards the application of the nested soft-collinear subtraction
scheme to the description of generic LHC processes by considering a situation when the hard emittors are in the
final state.  An important physical example of this situation is decays of colorless particles 
into a $q \bar q$ or $gg$ final state.  The NNLO QCD results for  the $q \bar q$
final state have already been used by us in Ref.~\cite{Caola:2017xuq} to describe the decay of the Higgs boson
into a massless $b \bar b $ pair; however,  we did not provide analytic formulas for this final state
in that reference.  The goal of this paper is to provide such formulas and to supplement them
with the analytic results for  decays of a color singlet into a   $gg$ final state.

Although, conceptually, the computation of NNLO QCD corrections to the production and  decay of a color
singlet are very similar, there are a few differences between the two that are worth pointing out.

\begin{itemize}

 \item In the case of the double-real corrections to the $gg$ final state we need to carefully
   separate  unresolved gluons from the resolved ones.  This issue does not appear in case of production
   where incoming particles are always the hardest ones and their momenta are fixed. 

\item  The computation of the integrated collinear counter-terms requires  modifications since, in the initial-state case,
  the integrated collinear subtraction terms are functions of fractions of the initial energy that a hard parton carries into the hard
  process,  while  in case of the final-state emissions one has to integrate over fractions of energies that are shared
  by partons in the collinear splitting. 

\item Construction of the double-collinear  phase space, i.e. the phase space appropriate for the description
  of a kinematic situation where singularities occur  when each unresolved parton is emitted by a different emittor, 
  is straightforward in the production and non-trivial  in the decay cases.

\item Obviously, no renormalization of  parton distribution functions is needed to describe  decay 
processes; for this
  reason,  cancellation of infra-red and
  collinear singularities  works differently in the production and decay cases. 

  \end{itemize} 

The rest of the paper is organized as follows.
In Section \ref{sect:2} we set the stage for the calculations described in the following
sections and introduce our notation.  We then discuss in detail the calculation of QCD corrections
to  $H \to gg$ decay to explain our approach. In particular, 
in Section \ref{sec:nlo},  we present    the computation of the NLO QCD corrections to the decay rate  $H \to gg$. 
In Section~\ref{sec:nnlogg} we discuss how to set up the calculation of NNLO QCD corrections to $H \to gg$ decay
and then consider  the $H \to 4g$ channel in detail.  We  present our final
results for the  NNLO QCD corrections to the decay of a color singlet  to two gluons in
Section~\ref{sec:nnlogg} and  to a $ q \bar q$ final state in Section~\ref{sec::nnloqq}.
We discuss the validation of our results in Section~\ref{sec::valid}, and  conclude in Section~\ref{sec::concl}.
Many useful formulas and intermediate results
are collected  in several appendices. 

\section{General considerations}
\label{sect:2}
We begin  by describing common features of QCD  corrections to color singlet decays and
by introducing notations that we will use throughout the paper. 
We  consider  decays of a color-singlet particle $Q$ to quarks and gluons. Our goal
is to provide formulas that describe  NNLO QCD corrections to these decays at a fully-differential  level. 
Specifically, we study  the decay process 
$Q \to f_i f_j + X$, where  $\{f_i,f_j\}$ can be either $\{g,g\}$ or $\{q,\qb\}$.  We first discuss
the decays into the  $gg$ final state since, compared to $Q \to q \bar q$,
the  singularity structure of the  decay $Q \to gg$   is more complex. Therefore,  
once the calculation of the NNLO QCD corrections to $Q \to gg$ is  understood, NNLO QCD
corrections to $Q \to q \bar q$   are easily established. 

We write the perturbative expansion of the differential decay rate as
\be
\d\Gamma = \d\Gamma^{\LO} + \d\Gamma^{\NLO} + \d\Gamma^{\NNLO} + ...
\label{eq2.1}
\ee
The different contributions in Eq.~(\ref{eq2.1}) are obtained by integrating various matrix elements
squared over the phase space of final state particles.  To describe this integration in a compact way, 
we introduce the notation analogous 
to our earlier papers~\cite{Caola:2017dug,Caola:2019nzf} and define 
\bes
\left\langle \FLM(1_{f_1},2_{f_2},...,n_{f_n}) {\cal O}(1,..,n) \right\rangle 
\equiv 
\frac{\mathcal N}{2 \mq}
\int \prod_{i=1}^{n} \df{i}
(2\pi)^d \delta^{(d)}(p_Q - p_1-p_2-...-p_n)
\\
\times |\mathcal M^{\rm tree}|^2(1_{f_1},2_{f_2},...,n_{f_n})
\mathcal O(\{p_1,...,p_n\}),
\end{split}
\ee
where $\mathcal N$ is a symmetry factor for identical final-state particles, $d=4-2\ep$ is the space-time dimensionality, 
\be
\df{i} = \frac{\d^{d-1}p_i}{(2\pi)^{d-1}2 E_i} \theta (\Em-E_i)
\label{eq2.3}
\ee
is the phase-space element for a parton $f_i$,
$\mathcal M^{\rm tree}(1_{f_1},...,n_{f_n})$ is the matrix element for
the process
\be
Q \to f_{1}(p_1)+f_2(p_2) + ... + f_n(p_n),
\ee
and $\mathcal O$ is a function that depends on partons' energies and angles. Furthermore, 
$\Em$ is an auxiliary  parameter with the dimension of energy that should be
large enough to accommodate all events that are allowed by the energy-momentum conservation
constraints.  Its  relevance will become clear in what follows.
In the rest of this paper, we will use $\Em = \mq/2$.
We note that the explicit constraint on the energy in Eq.~(\ref{eq2.3}) breaks Lorentz invariance
at intermediate stages  of the calculation; for 
this reason all energies in this paper are defined in the rest frame of the decaying particle $Q$. 

To describe contributions of  loop-corrected  processes, we introduce  similar quantities\footnote{
  We note that in this paper we always  work with UV-renormalized amplitudes.}   
\bes
\left\langle \FLV(1_{f_1},2_{f_2},...,n_{f_n}) {\cal O}(1,...,n)  \right\rangle 
\equiv 
\frac{\mathcal N}{2 \mq}
\int \prod_{i=1}^{n} \df{i}
(2\pi)^d \delta^{(d)}(p_Q - p_1-p_2-...-p_n)
\\
\times 2{\Re}\big[\mathcal M^{\rm tree} \mathcal M^{{\rm 1-loop},*}\big]
(1_{f_1},2_{f_2},...,n_{f_n})
\mathcal O(\{p_1,...,p_n\}),
\end{split}
\ee
and
\bes
\left\langle \FLVV(1_{f_1},2_{f_2},...,n_{f_n} {\cal O}(1,..,n) )\right\rangle 
\equiv 
\frac{\mathcal N}{2 \mq}
\int \prod_{i=1}^{n} \df{i}
(2\pi)^d \delta^{(d)}(p_Q - p_1-p_2...-p_n)
\\
\times \bigg[
2{\Re}\big[\mathcal M^{\rm tree} \mathcal M^{{\rm 2-loop},*}\big]
+\big|\mathcal M^{\rm 1-loop}\big|^2\bigg]
(1_{f_1},2_{f_2},...,n_{f_n})
\mathcal O(\{p_1,...,p_n\}).
\end{split}
\ee
Finally, we define
\be
\left\langle F_{X}(1,2,..
.,n) {\cal O}(1,...,n) \right\rangle = \sum_{f_1,f_2,...,f_n}
\left\langle F_{X}(1_{f_1},2_{f_2},...,n_{f_n}) {\cal O}(1,...,n) \right\rangle,
\ee
where  $X={\rm LM,~LV,~LVV}$ and   the sum runs over all allowed final states. 
Using these notations,  the three  contributions to the
differential width Eq.~(\ref{eq2.1}) are  written as 
\bes
& \d\Gamma^\LO = \left\langle \FLM(1,2)\right\rangle_\delta,\\
& \d\Gamma^\NLO = \left\langle \FLM(1,2,3)\right\rangle_\delta
+ \left\langle \FLV(1,2)\right\rangle_\delta,\\
& \d\Gamma^\NNLO = \left\langle \FLM(1,2,3,4)\right\rangle_\delta
+ \left\langle \FLV(1,2,3)\right\rangle_\delta
+ \left\langle \FLVV(1,2)\right\rangle_\delta.
\end{split}
\label{eq:LO...}
\ee
The symbol $\langle ... \rangle_\delta$ indicates that the
integration over the momenta of partons that are explicitly shown as arguments
of a function $F_X$ is not  performed,  so that the right hand side of Eq.~(\ref{eq:LO...})
provides a fully-differential description
of the decay rate. 

Starting from next-to-leading order,  the individual terms appearing on the right hand  sides 
of Eq.~\eqref{eq:LO...} are infra-red divergent and cannot be integrated in four dimensions when taken
separately.  The goal of a subtraction scheme is 
to rearrange them in the following way 
\bes
\d\Gamma^{\NLO} &= \d\Gamma^{\rm NLO}_{Q\to 2} + \d\Gamma^{\rm NLO}_{Q \to 3},\\
\d\Gamma^{\NNLO} &= \d\Gamma^{\rm NNLO}_{Q \to 2}  + \d\Gamma^{\rm NNLO}_{Q \to 3 }+\d\Gamma^{\rm NNLO}_{Q \to 4},
\end{split}
\label{eq:210}
\ee
where  $\d\Gamma^{\rm (N)NLO}_{Q \to i}$ are  finite in four dimensions and contain
contributions from final states with at most $i$ partons. 
In Refs.~\cite{Caola:2017dug,Caola:2019nzf} we explained how this can be done 
for hadroproduction of color-singlet states. We now use a very similar procedure to 
discuss color singlet decays. 

Since the required computations are often quite similar, we do not describe the 
calculational  details if the results  for the decay
follow easily from the ones for the production. To this end, we  note that  a detailed  introduction to our subtraction
scheme can be found in Refs.~\cite{Caola:2017dug,Caola:2019nzf} and we extensively refer to
these papers in what follows. In this paper, we highlight 
differences between the computations required for the production and decay cases
  and  present 
formulas for color singlet decay to either $gg$ or $q \bar q$ final states.
We begin with the  discussion of  the NLO QCD corrections to $H \to gg$.

\section{Higgs decay to gluons: a NLO computation}
\label{sec:nlo}
We consider the NLO QCD contribution to the differential decay rate of the Higgs boson to two gluons, $H \to gg$.\footnote{In this section and in Sec.~\ref{sec:nnlogg}, we assume that the Higgs directly 
couples to gluons through an effective vertex.}
We use the notations introduced in the previous section to write  
\be
\d\Gamma^{\NLO} = 
\left\langle \FLM(1_g,2_g,3_g)\right\rangle_\delta
+ \nf \left\langle \FLM(1_q,2_g,3_\qb)\right\rangle_\delta
+ \left\langle\FLV(1_g,2_g)\right\rangle_\delta,
\label{eq:212}
\ee
where $\nf$ is the number of massless quarks. 
We consider the three terms in Eq.~(\ref{eq:212}) separately, starting with  the real-emission contribution 
$\FLM(1_g,2_g,3_g)$.  The first step is to identify all possible singularities  that may appear in the computation of that contribution 
and to partition the phase space in such a way that for each  partition only a small subset of singularities is present. 

An important consequence of any  partitioning is the fact that certain partons are identified as ``hard''. This means that, for a given
partition, we should know exactly which partons cannot produce infra-red singularities. 
Although there are many ways to construct partitions, we find it  convenient to
use scalar products of the gluons' four-momenta  $s_{ij} = 2p_i \cdot p_j$ and the energy-momentum conservation
\be
p_H^2 = (p_1 + p_2 + p_3)^2 \;\;\; \Rightarrow \;\;\;\; \mq^2 = s_{12} + s_{13} + s_{23}, 
\ee
inside $\la\FLM(1_g,2_g,3_g)\ra_\delta$.  We then  write 
\bes
&\la\FLM(1_g,2_g,3_g)\ra_\delta = 
\\
&\quad\quad\quad\quad
\la\stij{12}\FLM(1_g,2_g,3_g)\ra_\delta + 
\la\stij{13}\FLM(1_g,2_g,3_g)\ra_\delta +
\la\stij{23}\FLM(1_g,2_g,3_g)\ra_\delta,  
\label{eq:part}
\end{split}
\ee
where we have introduced the notation  $\stij{ij}  \equiv s_{ij}/\mq^2$. We can use the symmetry of the
matrix element and the phase space to rewrite this equation as
\be
\la\FLM(1_g,2_g,3_g)\ra_\delta = 3 \la\st \FLM(1_g,2_g,3_g)\ra_\delta. 
\label{eq2.13}
\ee
Thanks to the prefactor $\tilde s_{12}$, gluons $g_1$ and $g_2$ on the right-hand side of Eq.~(\ref{eq2.13}) 
must be hard or ``resolved'' and the only potentially
unresolved parton is the gluon $g_3$.  This means that the right-hand side of Eq.~\eqref{eq2.13} is singular when
$g_3$ is soft and when $g_3$ is collinear to either $g_1$ or $g_2$; it is, however, not singular when either $g_1$ or $g_2$
is soft or when $g_1$ and $g_2$ are collinear to each other. 

We can follow the  approach described in the context of color singlet 
production~\cite{Caola:2017dug,Caola:2019nzf}
to extract singularities from the right hand side  of Eq.~(\ref{eq2.13}).  We begin by considering
the soft contribution  that arises when  energy of the gluon $g_3$,  $E_3$, becomes
small.  We find
\be
\lim_{E_3 \to 0} |{\cal M}^{\rm tree}|^2(1_g,2_g,3_g) \approx 
 2 C_A \gsb^2  \frac{p_1\cdot p_2}{(p_1\cdot p_3)(p_2\cdot p_3)}
|\mathcal M^{\rm tree}|^2(1_g,2_g), 
\label{eq:soft}
\ee
where $C_A = N_c=3$ is the $SU(3)$ color factor and $\gsb$ is the bare
strong coupling. 

The factorization formula Eq.~\eqref{eq:soft} allows us to  extract contributions of  soft
singularities  from  the decay rate.  To do so, we 
introduce the soft operator $S_3$ that extracts the most singular contributions in the soft limit from the
matrix element squared and the relevant phase space:
\bes
& \la S_3 \st \FLM(1_g,2_g,3_g)\ra_\delta = 
\frac{1}{2\mq}\frac{1}{3!}
\int \df1\df2 (2\pi)^d \delta^{(d)}(p_Q-p_1-p_2) 
|\mathcal M^{\rm tree}|^2(1_g,2_g) \\
& \times (2 C_A \gsb^2)\int \frac{d^{d-1} p_3}{(2\pi)^{d-1}2E_3} \theta(\Em-E_3)
 \frac{p_1\cdot p_2}{(p_1\cdot p_3)(p_2\cdot p_3)},
\end{split}
\label{eq:218}
\ee
Note that the function $\theta(E_{\rm max} - E_3)$ prevents the integral over $E_3$ from becoming unbounded from above. 
We rewrite Eq.~(\ref{eq:218}) as 
\be
\la S_3 \st \FLM(1_g,2_g,3_g)\ra_\delta = \frac{1}{3} \; \la    \la S_3\ra  \; \FLM(1_g,2_g)\   \ra_\delta, 
\ee
where we defined\footnote{We remind the reader that throughout this paper
we will use $\Em = \mq/2$.}
\bes
\la S_3\ra  \equiv (2 C_A \gsb^2)\int \frac{d^{d-1} p_3}{(2\pi)^{d-1}2E_3} \theta(\Em-E_3)
 \frac{p_1\cdot p_2}{(p_1\cdot p_3)(p_2 \cdot p_3)}\\
=
\frac{2 C_A \qas}{\ep^2}\lp\frac{\mq^2}{\mu^2}\rp^{-\ep}
(\eta_{12})^{-\ep}\left[
1 + \ep^2 \big[\Li_2(1-\eta_{12})-\zeta_2\big] + \mathcal O(\ep^3) \right],
\end{split}
\label{eq:219} 
\ee
together with
\be
\qas = \frac{\as(\mu)}{2\pi}\frac{e^{\ep\gamma_E}}{\Gamma(1-\ep)},
\ee
and
\be
\eta_{ij} = \frac{1-\cos \theta_{ij}}{2}.
\ee
We note that in the $H \to gg$ decay discussed here
$\eta_{12} = 1$; however, we do not use this fact right away and write 
Eq.~(\ref{eq:219}) in a  more general way.
The calculation that we just described allows us to remove the soft singularity. We obtain 
\be
3 \la \st \FLM(1_g,2_g,3_g)\ra_\delta = 
\la \la S_3\ra   \FLM(1_g,2_g)\ra_\delta + 
3 \la (\I-S_3)\st \FLM(1_g,2_g,3_g)\ra_\delta.
\label{eq:310}
\ee
We note that, since the reduced matrix
element does not require further regularization,   all singularities
in the first term on the r.h.s. of  Eq.~(\ref{eq:310}) are explicit.
The second term there is free
of soft singularities, but it still contains collinear ones; these  occur when 
$\eta_{31}=(1-\cos\theta_{31})/2$ or $\eta_{32}=(1-\cos\theta_{32})/2$ vanish. To isolate these singularities, we partition the 
phase space in such a way that only one of them can occur  at a time. To this end, we
introduce the partition of unity 
\be
1 = \w^{31} + \w^{32},
\label{eq:partu}
\ee
such that
$
\la \w^{31}(\I-S_3)\st \FLM(1_g,2_g,3_g)\ra
$
only has collinear singularities if $\eta_{31}\to 0$
and
$
\la \w^{32}(\I-S_3)\st \FLM(1_g,2_g,3_g)\ra
$
only has collinear singularities if $\eta_{32}\to 0$.
For example, one can choose\footnote{ Note that
this choice is always well-defined because the configuration 
$p_1||p_2||p_3$ is kinematically not allowed.}
\be
\w^{31} = 
\frac{\eta_{32}}{\eta_{31}+\eta_{32}},\;\;\;\;
\w^{32} = 
\frac{\eta_{31}}{\eta_{31}+\eta_{32}}.
\ee
Introducing this angular partitioning, we write 
\be
  \la (\I-S_3)\st \FLM(1_g,2_g,3_g)\ra_\delta   = \sum \limits_{i=1}^{2} 
  \la \w^{3i}(\I-S_3)\st \FLM(1_g,2_g,3_g)\ra_\delta,
\ee
and consider the two terms in the sum separately. 
We start with  the $i=1$ term. 
Similarly to the soft case,
we introduce a $C_{31}$ operator which  extracts the  corresponding  collinear singularity,
and apply it to $\la \omega^{31}(\I-S_3)\st\FLM(1_g,2_g,3_g)\ra$. We define 
$C_{31}$ in such a way that it extracts the leading $\eta_{31
} \to 0$ singularity from $\la \FLM(...) \ra_\delta $ without 
 acting  on the phase-space elements
 $[df_{1,.,3}]$, see Ref.~\cite{Caola:2019nzf} for more details. We find 
\bes
\la C_{31}\w^{31}\st \FLM(1_g,2_g,3_g)\ra_\delta 
&
\equiv
 \frac{1}{2 \mq}\frac{1}{3!}
\int \df1\df2\df3  
(2\pi)^d \delta^{(d)}(p_Q-p_2-p_{13}) 
\\
&\times 
\lp\frac{E_1}{E_1+E_3}\rp
\frac{\gsb^2}{p_1\cdot p_3} P_{gg}\lp \frac{E_1}{E_{13}}\rp \otimes
|\mathcal M^{\rm tree}|^2(13_g,2_g).
\end{split}
\label{eq:225}
\ee
In Eq.~\eqref{eq:225}, we defined 
\be
p_{13} \equiv \frac{E_{13}}{E_1}p_1,~~~ E_{13} = E_{1}+E_3, 
\ee
and denoted an on-shell  gluon with momentum $p_{13}$ as $13_g$. 
The function $P_{gg}$ in Eq.~\eqref{eq:225} stands for the $g^* \to gg$ splitting function and  
we used the  $\otimes$-sign to indicate  its  spin-correlated product 
with the matrix element squared, see Refs.~\cite{Caola:2017dug,Caola:2019nzf} for details.
In these references, we explicitly showed that at NLO
spin correlations disappear  after azimuthal averaging.
As the result, Eq.~(\ref{eq:225}) becomes 
\bes
&  \la C_{31}\w^{31}\st \FLM(1_g,2_g,3_g)\ra_\delta =
\\
&\quad\quad\quad
  \frac{1}{2 \mq}\frac{1}{3!}
\int \df1\df2 \d E_3(2\pi)^d \delta^{(d)}(p_Q-p_2-p_{13})
 |\mathcal M^{\rm tree}|^2(13_g,2_g)
\\
& \quad\quad\quad
\times \left[-\frac{\qas}{\ep}\frac{\Gamma^2(1-\ep)}{\Gamma(1-2\ep)}
\frac{\lp 4 E_3^2/\mu^2\rp^{-\ep}}{E_{13}}
 \la P_{gg}\ra  \lp \frac{E_1}{E_{13}}\rp
\theta(\Em-E_3)
\right],
\end{split}
\label{eq:227}
\ee
where $\la P_{gg}\ra$ is the spin-averaged $g^* \to gg $ splitting 
function
\be
\la P_{gg}\ra(z) = 2\Ca \left[\frac{1-z}{z}+\frac{z}{1-z} + z(1-z)\right].
\ee
The term on the second line of Eq.~\eqref{eq:227} is very similar to 
$\la \FLM(1_g,2_g)\ra_\delta$. To make this similarity explicit, we change integration
variables from $E_1$ and $E_3$ to $E_{13}$ and $z = E_1/E_{13}$.
We obtain
\be
E_1 = z E_{13},~ E_3 = (1-z) E_{13} 
\Rightarrow 
\frac{\df1 E_3^{-2\ep} \d E_3}{E_{13}} = \df{13} z \big[z(1-z)\big]^{-2\ep} 
E_{13}^{-2\ep}\d z.
\ee
We also rename $f_{13}$ back to  $f_1$ and obtain
\bes
&  \la C_{31}\w^{31}\st \FLM(1_g,2_g,3_g)\ra_\delta 
 =
 \\
& \frac{1}{3} \la \FLM(1_g,2_g)  \times
\left[-\frac{\qas}{\ep}\frac{\Gamma^2(1-\ep)}{\Gamma(1-2\ep)}
\lp\frac{4 E_{1}^2}{\mu^2}\rp^{-\ep}
\int\limits_{z_{\rm min}}^1 z \big[z(1-z)\big]^{-2\ep}\la P_{gg}\ra(z)
\d z
\right] \ra_\delta,
\end{split}
\label{eq:231}
\ee
where $E_1 = \mq/2$ and we used the fact that the integration over $z$ starts at 
$z=z_{\rm min} = {\rm min}\{0,1 - \Em/E_{1}\}$. 
Since $\Em$ must be chosen in  such a way that the whole phase space is covered,
$E_{\rm max}$ should be larger than $E_1$, $E_{\rm max} > E_1$,  for all $E_1$. This implies  $\zm = 0$. 

Repeating these steps for the soft-collinear term $S_3 C_{31}$,  we find  
\bes
 \la S_3 C_{31}\w^{31}\st \FLM(1_g,2_g,3_g)\ra_\delta  &= \frac{1}{3} \la \FLM(1_g,2_g)\ra_\delta 
\\
& \times 
\left[-\frac{\qas}{\ep}\frac{\Gamma^2(1-\ep)}{\Gamma(1-2\ep)}
\lp\frac{\mq^2}{\mu^2}\rp^{-\ep}
\int\limits_{0}^1  \frac{2\Ca}{(1-z)^{1+2\ep}}
\d z
\right].
\end{split}
\label{eq:232}
\ee
We use these results to write 
\bes
& \la w^{31}(\I-S_3) \st \FLM(1_g,2_g,3_g)\ra_\delta = \frac{1}{3} \la C_{31}(\I-S_3) \ra \times \la\FLM(1_g,2_g)\ra_\delta 
\\
& 
+ \la (I - C_{31}) w^{31}(\I-S_3) \st \FLM(1_g,2_g,3_g)\ra_\delta,
\end{split}
\label{eq:233}
\ee
where $\la C_{31}(\I-S_3) \ra  $ follows from Eqs.~(\ref{eq:231},\ref{eq:232}). We 
find 
\bes
\la C_{31}(\I-S_3)\ra  = \frac{\qas}{\ep}
\frac{\Gamma^2(1-\ep)}{\Gamma(1-2\ep)}
\lp\frac{\mq^2}{\mu^2}\rp^{-\ep}
 \gamma_{z,g \to gg}^{22}
\end{split}
\label{eq:new323}
\ee
where  $\gamma_{z,g \to gg}^{22}$ is a particular case of a general anomalous dimension defined as follows 
  \be
\gamma^{n k }_{f(z),g\to gg} 
= - \int\limits_0^1 \d z\bigg[z^{-n \ep}(1-z)^{-k \ep}
  \la f(z) P_{gg}(z)\ra - 2\Ca f(1) (1-z)^{-1-k\ep} \bigg].
\label{eq433a}
\ee

We note that in the first term on the right hand side  in  Eq.~\eqref{eq:233}
all singularities are manifest and the reduced matrix element does not
require regularization, whereas  the second term is free of \emph{both}
soft and collinear singularities so that it can be immediately  integrated in four
dimensions. 

We deal with the  $\w^{32}$ term in the  partition of unity
Eq.~(\ref{eq:partu}) in a similar way. We obtain 
\bes
& \la w^{32}(\I-S_3) \st \FLM(1_g,2_g,3_g)\ra_\delta = \frac{1}{3} \la C_{32}(\I-S_3) \ra  \times \la \FLM(1_g,2_g)\ra_\delta \\
& + \la (I-C_{32}) w^{32}(\I-S_3) \st \FLM(1_g,2_g,3_g)\ra_\delta,
\end{split}
\label{eq:236}
\ee
with 
\bes
\la C_{32}(\I-S_3)\ra  = \frac{\qas}{\ep}
\frac{\Gamma^2(1-\ep)}{\Gamma(1-2\ep)}
 \lp\frac{\mq^2}{\mu^2}\rp^{-\ep} \gamma_{z,g \to gg}^{22} = \la C_{31}(\I-S_3)\ra.
 \end{split}
\ee

We combine Eqs.~(\ref{eq:310},\ref{eq:new323},\ref{eq:236}) and obtain the following result for the three-gluon  contribution
to Higgs boson decay 
\bes
\la\FLM(1_g,2_g,3_g)\ra_\delta = 
 \bigg[\la S_3\ra + 2 \la C_{31}(\I-S_3)\ra\bigg]
\times \la  \FLM(1_g,2_g)\ra_\delta 
\\
+  3 \sum_{i=1,2} \la (\I-C_{3i})\w^{3i}(\I-S_3) \st \FLM(1_g,2_g,3_g)\ra_\delta.
\label{eq:237}
\end{split}
\ee
We note that, thanks to Bose symmetry, the two terms in the sum in the last
line in Eq.~(\ref{eq:237}) are the same. Hence, we write
\bes
\la\FLM(1_g,2_g,3_g)\ra_\delta = 
\bigg[\la S_3\ra + 2 \la C_{31}(1-S_3)\ra\bigg]
\times \la  \FLM(1_g,2_g)\ra_\delta
\\
 +  6 \la (\I-C_{31})\w^{31}(\I-S_3) \st \FLM(1_g,2_g,3_g)\ra_\delta.
\end{split}
\label{eq:238}
\ee
This discussion implies that   Bose symmetry can be efficiently used
to partition the phase space in
such a way that identical kinematic configurations of the three-gluon  final states are accounted
for only once in the calculation; this removes  the original $1/3!$ symmetry factor. 

Before combining this result with virtual corrections, we consider the other
real-emission term in Eq.~\eqref{eq:212},  $\nf \la \FLM(1_q,2_g,3_\qb )\ra$, that
describes the decay $H \to  (g^* \to q \bar q)  g$. Because (in this section) the $q \bar q$ pair
does not directly couple to the Higgs boson, the singularity in this case  is produced 
by the collinear splitting $g^*\to q\qb$.  For this reason, we do not need any partitioning. 
We repeat  steps that led to Eq.~\eqref{eq:233} and obtain\footnote{The extra factor of
2 comes from a mismatch between the symmetry factors of 
$\la \FLM (1_q,2_g,3_\qb) \ra$ and $\la\FLM(1_g,2_g) \ra$.}
\bes
\nf \la \FLM(1_q,2_g,3_\qb)\ra_\delta = 
\nf \la (\I-C_{31})\FLM(1_q,2_g,3_\qb)\ra_\delta
\\+ 
2\nf \frac{\qas}{\ep} \frac{\Gamma^2(1-\ep)}{\Gamma(1-2\ep)} 
\lp\frac{\mq^2}{\mu^2}\rp^{-\ep} \gamma_{1,g \to q \bar q}^{22} \;  \la \FLM(1_g,2_g)\ra_\delta,
\end{split}
\label{eq:240}
\ee
where 
\be
 \gamma_{1,g \to q \bar q}^{22}  = -\int\limits_0^1 \d z 
\big[z(1-z)\big]^{-2\ep} \la P_{gq}\ra(z),
~~~~ \la P_{gq}\ra (z) = \tr \left[ 1 - \frac{2z (1-z)}{1-\ep}\right].
\label{eq:241}
\ee

We can now combine the $H\to ggg$ and $H\to qg\qb$ decay channels  and write the 
total real-emission contribution to $\d\Gamma^\NLO$, up to higher orders in $\ep$, as
\bes
& \la\FLM(1,2,3)\ra_\delta = \la\FLM(1_g,2_g,3_g)\ra_\delta + \nf \la \FLM(1_q,2_g,3_\qb)\ra_\delta
=\qas   \lp\frac{\mq^2}{\mu^2}\rp^{-\ep} 
\\
&  \times\Bigg ( \frac{2 C_A}{\ep^2} 
\bigg[1+\ep^2\big[\Li_2(1-\eta_{12})-\zeta_2 \big] \bigg ]
   + \frac{2 \gamma_{g}(\ep)}{\ep} +\mathcal O(\ep)
  \Bigg ) \langle  \FLM(1_g,2_g)  \rangle_\delta 
\\
& + 6 \la (\I-C_{31})\w^{31}(\I-S_3) \st \FLM(1_g,2_g,3_g)\ra_\delta  +
\nf\la (\I-C_{31})\FLM(1_q,2_g,3_\qb)\ra_\delta ,
\end{split}
\label{eq:244}
\ee
where we have defined
\be
\gamma_g(\ep) = \gamma^{22}_{z,g \to gg}  + \nf \gamma^{22}_{1,g \to q \bar q}(\ep) = 
\gamma_g + \ep \gamma_g' + {\cal O}(\ep^2).
\ee
The two quantities $\gamma_g$ and $\gamma_g'$ are given in Eq.~\eqref{eq:app_gamma_g}.

It remains to  combine Eq.~\eqref{eq:244} with virtual corrections. 
We follow Ref.~\cite{Catani:1998bh} to  separate the divergent and finite parts
of the one-loop amplitude and define
\bes
&
\la\FLV(1_g,2_g)\ra_\delta = \la\FLVfin(1_g,2_g)\ra_\delta
\\
& 
-2  \qas\cos(\ep\pi) C_A \lp\frac{1}{\ep^2} + \frac{\gamma_g}{C_{A}\ep}\rp \la 
 \lp\frac{4 E_1 E_2 \eta_{12}}{\mu^2}\rp^{-\ep}
\FLM(1_g,2_g) \ra_\delta,
\end{split}
\label{eq:245}
\ee
where  $\la\FLVfin(1_g,2_g)\ra_\delta$ is a finite remainder of the one-loop $H \to gg$ amplitude, 
see Appendix A in Ref.~\cite{Caola:2019nzf} for details. 
We combine Eq.~(\ref{eq:244}) and Eq.~(\ref{eq:245}), use $\eta_{12} = 1$ and obtain
a very simple result for the NLO QCD corrections to $H \to gg$ decay.
It reads 
\bes
& \d\Gamma^{\rm NLO}_{Q \to 2}   = \asontwopimu
\lp 2 \gamma'_g +\frac{2\pi^2}{3}\Ca\rp\la\FLM(1_g,2_g)\ra_\delta
+\la \FLVfin(1_g,2_g)\ra_\delta,
\\
& \d\Gamma^\NLO_{Q\to 3}  = 
\la (\I-C_{31}) ( 6 \w^{31}(\I-S_3) \st \FLM(1_g,2_g,3_g)
 + n_f \FLMs(1_q,2_g,3_\qb) \right ) \rangle_\delta, 
\end{split}
\label{eq:248}
\ee
where the two contributions are defined  in 
Eq.~(\ref{eq:210}). 

We conclude this section by reminding the reader 
that the NLO construction we just described is identical to the
FKS subtraction scheme~\cite{Frixione:1995ms,Frixione:1997np}. In the next sections, we will
show how to generalize the FKS scheme to NNLO.

\section{Higgs decay to gluons: a NNLO computation}
\label{sec:nnlogg}

In this section we generalize the discussion of the NLO QCD corrections to the 
decay of a color singlet
 to the NNLO case. We will follow 
Refs.~\cite{Caola:2017dug,Caola:2019nzf}  and  perform subtractions of  soft and collinear 
divergences in an iterated manner, starting from the soft ones. 
Many technical details are similar  to the production case
described at length in the above references and we do not discuss them here. 
Instead, we  focus on the peculiarities  of the decay.

\subsection{Double-real contribution}

There are four different partonic final states that we have to consider. They are
{\it a}) 4 gluons, {\it b})  2 gluons, 2 quarks, {\it c}) two quark pairs of different flavors
and {\it d}) two quark pairs of the same flavor.  We write 
\bes
\la \FLM(1,2,3,4)\ra_\delta & = 
\la\FLM(1_g,2_g,3_g,4_g)\ra_\delta+
\nf \la\FLM(1_g,2_g,3_q,4_\qb)\ra_\delta
\\
& + \frac{\nf(\nf-1)}{2}\la\FLM(1_q,2_{q'},3_\qb,4_{\qb'})\ra_\delta
+\nf\la\FLM(1_q,2_q,3_\qb,4_\qb)\ra_\delta.
\end{split}
\ee
In full analogy  to the NLO case,  we  partition  the phase space in such a way that only a
subset of partons are allowed to become unresolved. In case of the NNLO contributions, {\it two}
partons can become unresolved simultaneously; we will systematically rename partons so that,
eventually, the unresolved  partons are  always referred to as $f_3$ and $f_4$. 

We first consider the four-gluon channel,
\be
H \to g(p_1) g(p_2) g(p_3) g(p_4),
\ee
and introduce a  partition of unity following  what has already been done at NLO
\be
1=\tilde{s}_{12}+\tilde{s}_{13}+\tilde{s}_{14}+\tilde{s}_{23}+\tilde{s}_{24}+\tilde{s}_{34}.
\label{eq:4.4}
\ee
We insert this partition inside the integrand for $\la \FLM(1_g,2_g,3_g,4_g)\ra_\delta $, 
use  the symmetry of the phase space and the matrix element
and arrive at\footnote{In this subsection, the ``tree'' superscript on $\mathcal M$
is always assumed.}
\bes
&2\mq\la \FLM(1_g,2_g,3_g,4_g)\ra =
\\
&\quad\quad\quad\quad\quad\quad
\frac{1}{4!} \int \prod \limits_{i=1}^{4} [df_i] \;(2\pi)^d \delta^{(d)}( p_Q - \sum \limits_{i=1}^{4} p_i)
\sum \limits_{i \ne j=1}^{4} \tilde{s}_{ij} |{\cal M}(1_g,2_g,3_g,4_g)|^2=\\
&\quad\quad\quad\quad\quad\quad
\frac{1}{4} \int  \prod \limits_{i=1}^{4} [df_i] \;(2\pi)^d \delta^{(d)}(p_Q - \sum \limits_{i=1}^{4} p_i) \; \tilde{s}_{12} \; |{\cal M}(1_g,2_g,3_g,4_g)|^2.
\end{split}
\ee
The  prefactor ${\tilde s}_{12}$ ensures  that no singularity arises
in the product $\tilde{s}_{12} \; |M(1_g,2_g,3_g,4_g)|^2$ when
gluons $1$ and $2$ become either soft or collinear to each other.
To proceed further, we introduce an energy ordering for potentially-unresolved
gluons $g_3$ and $g_4$, use $g_3 \leftrightarrow g_4$ symmetry  and write 
\bes
& 2\mq\la \FLM(1_g,2_g,3_g,4_g)\ra 
 =\frac{1}{2}  \int  \prod \limits_{i=1}^{4} [df_i] \;(2\pi)^d \delta^{(d)}(p_Q - \sum \limits_{i=1}^{4} p_i)\tilde{s}_{12}  \theta(E_3-E_4) \times
 \\
& \quad\quad\quad\quad
 |{\cal M}(1_g,2_g,3_g,4_g)|^2 = 12\times(2 m_H^2)
 \left\langle \st \FLM(1_g,2_g,3_g,4_g) \theta(E_3 - E_4)\right \rangle.
\end{split} 
\ee

We now consider the $2q2g$ final state. In principle,
it contains fewer singularities than the four-gluon  final state. Therefore,
one may use a simpler partition of
unity to single out the potentially unresolved partons. However, to
streamline the bookkeeping,  we find it convenient
to use identical partitioning for all  final states. Our starting point is then
\bes
&2\mq\la \FLM(1_g,2_g,3_q,4_\qb) \ra =
\\\
&\quad\quad\quad\quad
\frac{1}{2!} 
\int  \prod \limits_{i=1}^{4} [df_i] \;(2\pi)^d \delta^{(d)}(p_Q - \sum \limits_{i=1}^{4} p_i)
\sum \limits_{i \ne j =1}^{4} \tilde{s}_{ij} 
|{\cal M}(1_g,2_g,3_q,4_{\bar q})|^2,
\end{split}
\label{eq2.33}
\ee
where the partition of unity Eq.~(\ref{eq:4.4}) has already been employed. We note that the
amplitude is symmetric with respect to permutations of the two gluons,  so that
\be
|{\cal M}(i_g,j_g, k_q,l_\qb)|^2 = |{\cal M}(j_g,i_g, k_q,l_\qb)|^2.
\ee 
Furthermore, since in this  amplitude the quark-antiquark pair arises from gluon 
splitting, the amplitude squared summed over quark and anti-quark polarizations
satisfies $|{\cal M}(i_g,j_g,k_q,l_\qb )|^2=|{\cal M}(i_g,j_g,l_q,k_\qb)|^2$.
We can use these symmetries of the amplitude squared
as well as the symmetry  of the phase space to re-write Eq.~(\ref{eq2.33}) in the following way
\bes
&2\mq \la \FLM(1_g,2_g,3_q,4_\qb) \ra 
 =
 \frac{1}{2!} \int  \prod \limits_{i=1}^{4} [df_i] \;(2\pi)^d \delta^{(d)}(p_Q - \sum \limits_{i=1}^{4} p_i)
 \times \; 
\\
&\;\;\;\;\;\;\;
\tilde{s}_{12} \biggl( |{\cal M}(1_g,2_g,3_q,4_\qb)|^2
+|{\cal M}(1_g,3_g,2_q,4_\qb)|^2
+|{\cal M}(1_g,4_g,2_q,3_\qb)|^2
\\
&\;\;\;\;\;\; +|{\cal M}(2_g,3_g,1_q,4_\qb)|^2 +|{\cal M}(2_g,4_g,1_q,3_\qb)|^2 +|{\cal M}(3_g,4_g,1_q,2_\qb)|^2 \biggr).
  \end{split}
\label{eq2.36}
\ee
To proceed further, we introduce the energy ordering for the two potentially unresolved partons $f_{3,4}$ and use symmetries of the amplitude
to remove the factor $1/2$ in the above equation. In cases when $f_{3}$ and $f_{4}$ are partons of a different type,
this requires us to combine the different contributions in a particular way. As an example, consider  the second and  the third
term in Eq.~(\ref{eq2.36}). Relabelling parton momenta where appropriate, we write
\be
\begin{split} 
  & |{\cal M}(1_g,3_g,2_q,4_\qb)|^2 +|{\cal M}(1_g,4_g,2_q,3_\qb)|^2  =
  \big[ |{\cal M}(1_g,3_g,2_q,4_\qb)|^2 + |{\cal M}(1_g,4_g,2_q,3_\qb)|^2  
  \\
&\;\;\;\;\;\;\;\;\;\;\;\;\;\;\;\;\;\;\;\;\;\;\;\;\;\;\;\;
+ |{\cal M}(1_g,4_g,2_q,3_\qb)|^2   + |{\cal M}(1_g,3_g,2_q,4_\qb)|^2 \big ] \theta(E_3 - E_4)
\\
& \;\;\;\;\;\;\;\;\;\;\;\;\;\;\;\;\;\;\;\;\;\;\;\;\;\;\;\; = 2 \left [ |{\cal M}(1_g,3_g,2_q,4_\qb)|^2 + |{\cal M}(1_g,4_g,2_q,3_\qb)|^2 \right ] \theta(E_3 - E_4).
\end{split} 
\ee
Using these transformations, we obtain 
\bes
& 2\mq\la \FLM(1_g,2_g,3_q,4_\qb) \ra  
= \int  \prod \limits_{i=1}^{4} [df_i] \;(2\pi)^d \delta^{(d)}(p_Q - \sum \limits_{i=1}^{4} p_i)
\; \theta(E_3 - E_4)
\\
& \;\;\;\;\;\;\;\;
\times \tilde{s}_{12} \bigl( |{\cal M}(1_g,2_g, 3_q,4_\qb)|^2
+|{\cal M}(1_g,3_g,2_q,4_\qb)|^2 +|{\cal M}(1_g,4_g,2_q,3_\qb)|^2
\\
& \;\;\;\;\;\;\;\;
+|{\cal M}(2_g,3_g,1_q,4_\qb)|^2
+|{\cal M}(2_g,4_g,1_q,3_\qb)|^2
+|{\cal M}(3_g,4_g,1_q,2_\qb)|^2 \bigr),
  \end{split}
\ee
which we can write as
\bes
& \la \FLM(1_g,2_g,3_q,4_\qb) \ra   =
2  \bigg \langle  \bigg [ \FLM(1_g,2_g,3_q,4_\qb) 
  + \FLM(1_g,3_g,2_q,4_\qb) 
  \\
    &\;\;\;\;\;
+ \FLM(1_g,4_g,2_q,3_\qb) 
+ \FLM(2_g,3_g,1_q,4_\qb) 
+ \FLM(2_g,4_g,1_q,3_\qb)  
\\
&\;\;\;\;\;
+  \FLM(3_g,4_g,1_q,2_\qb)   \bigg ]\st \theta(E_3-E_4)  \bigg \rangle.
\label{eq2.39}
\end{split}
\ee
We note that the six terms in Eq.~(\ref{eq2.39}) have very different singularity structures.
For example, all the terms in Eq.~(\ref{eq2.39}) that contain gluon $g_4$ 
give  rise to single soft singularities that arise when $E_4 \to 0$. 
In the remaining three terms, the energy $E_4$ is associated with an anti-quark and, therefore,
these terms are not singular in the single-soft limit.
Similarly, the collinear limit $C_{41}$ corresponds to an (anti)quark and a gluon becoming collinear in the
first, second, fifth and sixth terms in Eq.~(\ref{eq2.39}). However, the same limit 
describes a kinematic configuration with  two collinear
gluons  in the third term in Eq.~(\ref{eq2.39}).  Clearly, the two limiting
cases result in  different splitting functions and different reduced matrix elements. 

Finally, we turn to the four-quark channels, where  we need to
make a further distinction between cases when quarks have same or different flavors. 
If they are different,  i.e. $q \ne q'$, we write  
\bes
2\mq&\left[\frac{\nf(\nf-1)}{2}\right] \la \FLM(1_q,2_{q'},3_\qb,4_{\qb'} ) \ra 
= 
\\
&
\frac{\nf(\nf-1)}{2} \int \prod \limits_{i=1}^{4} [df_i] \;(2\pi)^d \delta^{(d)}(p_Q - \sum \limits_{i=1}^{4} p_i)
\; |{\cal M}(1_q,2_{q'},3_{\qb},4_{\qb'})|^2.
\end{split}
\ee
If the flavors are identical, we can use the same amplitude ${\cal M}$ as for the different-flavor case,
accounting  for a permutation of two identical particles. We write 
\bes
2\mq \;\nf\; \langle \FLM(1_q,2_{q},3_\qb,4_{\qb}) \rangle
& = \frac{\nf}{(2!)^2}  \int \prod \limits_{i=1}^{4} [df_i] \;(2\pi)^d \delta^{(d)}(p_Q - \sum \limits_{i=1}^{4} p_i)
\\
& \times |{\cal M}(1_q,2_{q'},3_\qb,4_{\qb'}) - {\cal M}(1_q,2_{q'},4_\qb,3_{\qb'})|^2.
\end{split}
\ee
We denote the interference term as 
\be
   {\rm Int}(1_q,2_q,3_\qb,4_\qb) = -2{\rm Re}\left( {\cal M}(1_q,2_{q'},3_\qb,4_{\qb'})
   {\cal M}^*(1_q,2_{q'},4_\qb,3_{\qb'}) \right),
   \ee
   and write the {\it complete}  four-quark contribution to the decay rate, including both different and identical flavors,
   as 
\bes
2\mq\langle \FLM^{(4q)}(1,2,3,4) \rangle
& = \frac{\nf^2}{2} \int \prod \limits_{i=1}^{4} [df_i] \;(2\pi)^d \delta^{(d)}(p_Q - \sum \limits_{i=1}^{4} p_i)
 |{\cal M}(1_q,2_{q'},3_{\qb},4_{\qb'})|^2 \\
& + \frac{\nf}{4} \int \prod \limits_{i=1}^{4} [df_i] \;(2\pi)^d \delta^{(d)}(p_Q - \sum \limits_{i=1}^{4} p_i)  {\rm Int}(1,2,3,4).
\label{eq2.43}
\end{split}
\ee   
The interference term in Eq.~(\ref{eq2.43}) is not singular and can be evaluated in four dimensions;  for this reason we keep it as it is.
Moreover, the first term in that
equation only produces   singularities when either one or two $q\qb$ pairs become collinear.
Despite this simplicity, we find it convenient to treat the four-quark contributions Eq.~(\ref{eq2.43})
in the same way as the two other channels that we discussed previously. 
To this end,
we insert the partition of unity Eq.~(\ref{eq:4.4})
into the integrands in Eq.~(\ref{eq2.43}), re-label partonic momenta, use
the symmetry of the amplitude squared
\be
|{\cal M}(i_q,j_{q'},k_\qb,l_{\qb'})|^2 = |{\cal M}(i_q,l_{q'},k_\qb,j_{\qb'})|^2 =  |{\cal M}(k_q,j_{q'},i_\qb,l_{\qb'})|^2,
\ee
and obtain 
\bes
2\mq\langle \FLM^{(4q)}(1,2,3,4) \rangle  = &
\nf^2  \int \prod \limits_{i=1}^{4} [d f_i]
\;(2\pi)^d \delta^{(d)}(p_Q - \sum \limits_{i=1}^{4} p_i)
\\
& \; \times \st
\biggl( 2 |{\cal M}(1_q,2_{q'},3_\qb,4_{\qb'} )|^2+ |{\cal M}(1_q,3_{q'},2_{\qb},4_{\qb'})|^2 \biggr) \\
&+ \frac{\nf}{4}  \int \prod \limits_{i=1}^{4} [d f_i] \;(2\pi)^d \delta^{(d)}(p_Q - \sum \limits_{i=1}^{4} p_i) \; {\rm Int}(1,2,3,4). 
\end{split}
\ee

The prospective unresolved partons are $f_{3,4}$. Similar to other channels, we introduce the 
energy ordering $E_3 > E_4$ and again use the symmetry of the amplitude squared to simplify the result.
We obtain 
\bes
&\langle \FLM^{(4q)}(1,2,3,4) \rangle =  
\nf \la \FLM^{\rm int}(1_q,2_{q},3_\qb,4_{\qb}) \ra+
2\nf^2
 \big\langle
   \big[ \FLM(1_q,2_{q'},3_\qb,4_{\qb'}) + 
   \\
   &\quad\quad\quad\quad\quad\quad\quad
    \FLM(1_q,2_{q'},4_\qb,3_{\qb'})   
  + \FLM(1_q,3_{q'},2_\qb,4_{\qb'}) \big]
  \st \theta(E_3 - E_4) \big\rangle,
\end{split}
\ee   
where we have defined
\be
\nf\la \FLM^{\rm int}(1_q,2_{q},3_\qb,4_{\qb}) \ra = 
\frac{\nf}{4} \left[\frac{1}{2\mq}\right]
\int \prod \limits_{i=1}^{4} [d f_i] \;(2\pi)^d \delta^{(d)}(p_Q - \sum \limits_{i=1}^{4} p_i) \; {\rm Int}(1,2,3,4).
\ee

Upon combining all the  channels, we obtain the final result for the double-real contribution
to the decay width. It reads 
\bes
&
\la \FLM(1,2,3,4) \ra_\delta  = 
\Bigg\langle \st\theta(E_3 - E_4)\times\Bigg\{
12 \FLM(1_g,2_g,3_g,4_g)
\\
&+ 2\nf \bigg[
\FLM(1_g,2_g,3_q,4_\qb)
+\FLM(1_g,3_g,2_q,4_\qb)
+\FLM(1_g,4_g,2_q,3_\qb)
\\
&\quad\quad\quad+\FLM(2_g,3_g,1_q,4_\qb)
+\FLM(2_g,4_g,1_q,3_\qb)
+\FLM(3_g,4_g,1_q,2_\qb)
\bigg]
\\
&+2\nf^2 \bigg[
\FLM(1_q,2_{q'},3_\qb,4_{\qb'})+
\FLM(1_q,2_{q'},4_{\qb},3_{\qb'})+
\FLM(1_q,3_{q'},2_\qb,4_{\qb'})\bigg]
\Bigg\}
\\
&
+\nf\FLM^{\rm int}(1_q,2_q,3_\qb,4_\qb) 
\Bigg\rangle_\delta.
\end{split}
\label{eq:419a}
\ee

To illustrate how  soft and collinear singularities are extracted  from the double-real emission
contribution Eq.~(\ref{eq:419a}), we focus on the four-gluon final state $\FLM(1_g,2_g,3_g,4_g)$.
This contribution possesses the richest singularity structure yet, at the same time,  it is one of the simplest
as far as the  bookkeeping is concerned.  After explaining how the singularities are extracted
in this case,  we present the results for all channels in Section~\ref{sechggfull}.

\subsubsection{Double-soft contribution for $H \to gggg$ }

Similar to the production case, we begin with the double-soft limit that occurs when $E_3,E_4\to 0$. 
We follow  Refs.~\cite{Caola:2017dug,Caola:2019nzf} and  introduce an operator $\SS$ that extracts the leading double-soft
singularity from
the product of the matrix element squared 
and the phase space,   and write 
\be
\I = \SS + (\I-\SS).
\label{eq4.20a}
\ee
The double-soft limit is computed in exactly the same way as in the production case~\cite{Caola:2017dug,Caola:2019nzf}.
We find 
\bes
& 12\la \SS \st \FLM(1_g,2_g,3_g,4_g) \theta(E_3-E_4)\ra_\delta =
\qas^2 \Ca^2 D_{\SS}
  \lp\frac{\mq^2}{\mu^2}\rp^{-2\ep} \la \FLM(1_g,2_g)\ra_\delta,
\end{split}
\ee
where~\cite{max_soft}
\begin{align}
&   D_{\SS} = 
\frac{5}{2\ep^4} + \frac{11}{12\ep^3} 
+ \frac{1}{\ep^2} \left ( -\frac{16}{9} - \frac{11\pi^2}{12} + \frac{11}{3}\ln 2\right ) \nn \\
& + \frac{1}{\ep} \left ( \frac{217}{54} - \frac{11 \pi^2}{36} - 
\frac{137}{18}\ln 2
 - \frac{11}{3} \ln^2 2 
- \frac{53}{4}\zeta_3 \right ) \\
& -\frac{649}{81} + \frac{125\pi^2}{216} - \frac{131 \pi^4}{720} 
+\frac{434}{27}\ln 2 - \frac{11}{6}\pi^2 \ln 2
+\frac{137}{18}\ln^2 2 + \frac{22}{9} \ln^3 2 -\frac{275}{12}\zeta_3.
\nonumber 
\end{align}
We use Eq.~(\ref{eq4.20a}) and write 
\bes
&
12\la\st\FLM(1_g,2_g,3_g,4_g)\theta(E_3-E_4)\ra_\delta =
\qas^2 \lp\frac{\mq^2}{\mu^2}\rp^{-2\ep} \Ca^2 D_{\SS}
\la\FLM(1_g,2_g)\ra_\delta
\\
&
+12\la(\I-\SS)\st\FLM(1_g,2_g,3_g,4_g)\theta(E_3 - E_4)\ra_\delta.
\end{split}
\label{eq4.24}
\ee
The term on the second line in Eq.~(\ref{eq4.24}) 
does not contain double-soft singularities anymore 
but it still contains both single-soft and collinear ones. We discuss how to extract them
 in what follows.  

\subsubsection{Single-soft contribution}
We need to extract the single-soft singularity from 
\be
12\la(\I-\SS)\st\FLM(1_g,2_g,3_g,4_g)\theta(E_3 - E_4)\ra_\delta,
\ee
see Eq.~(\ref{eq4.24}).
The soft limit of the amplitude squared reads 
\be
S_4 |\mathcal M^{\rm tree}(1_g,2_g,3_g,4_g)|^2 = 
\gsb^2 C_A \sum \limits_{(ij) \in {1,2,3}}^{} \frac{p_i \cdot p_j}{(p_i \cdot p_4) ( p_j \cdot p_4)}
|\mathcal M^{\rm tree}(1_g,2_g,3_g)|^2,
\ee
where the sum runs over three $ij$-pairs,  $\{1,2\},\{1,3\},\{2,3\}$. 
The gluon $g_4$ decouples both from the hard matrix element and the phase-space; hence,
integration over its four-momentum is identical to the NLO case except that the 
upper  boundary for the  $E_4$  integration is now $E_3$. Repeating steps analogous
to what we discussed at NLO, we find
\bes
&12\la S_4(\I-\SS)\st \theta(E_3 - E_4)\FLM(1_g,2_g,3_g,4_g)\ra_\delta = 
\\
&\quad\quad\quad\quad\quad\quad\quad\quad\quad\quad
3\la J_{S_4}^{ggg} (\I-S_3)\st \FLM(1_g,2_g,3_g)\ra_\delta,
\end{split}
\label{eq:3.28}
\ee
where 
\bes
J_{S_4}^{ggg} = 
\qas\lp\frac{4 E_3^2}{\mu^2}\rp^{-\ep} \; \frac{\Ca}{\ep^2}\;
\Bigg[
\lp\eta_{12}\rp^{-\ep} K_{12} +
\lp\eta_{13}\rp^{-\ep} K_{13} +
\lp\eta_{23}\rp^{-\ep} K_{23}
\Bigg],
\end{split}
\ee
and 
\be
K_{ij} = \frac{\Gamma^2(1-\ep)}{\Gamma(1-2\ep)} 
\eta_{ij}^{1+\ep}~_2 F_1(1,1,1-\ep;1-\eta_{ij}) = 
1 + \ep^2\big[\Li_2(1-\eta_{ij}) - \zeta_2\big] + \mathcal O(\ep^3).
\ee

Eq.~\eqref{eq:3.28} is free from soft singularities, but it still contains collinear ones; these 
arise when the gluon $g_3$ becomes collinear to gluon $g_1$  or gluon $g_2$. 
We proceed as in the NLO computation. Namely, we introduce a partition of unity, 
use the symmetry of the process under the exchange of gluons 1 and 2 and write
\bes
& \la (\I-S_3) J^{ggg}_{S_4} \st \FLM(1_g,2_g,3_g) \ra_\delta = 
2 
\la C_{31} \w^{31} (\I-S_3) J^{ggg}_{S_4} 
\st \FLM(1_g,2_g,3_g) \ra_\delta
\\
&\quad\quad\quad\quad
+\la \sum_{i=1,2} (\I-C_{3i}) \w^{3i} (\I-S_3) J^{ggg}_{S_4} 
\st \FLM(1_g,2_g,3_g) \ra_\delta.
\end{split}
\label{eq:c31J}
\ee
All singularities are regulated in the second term on the r.h.s. of Eq.~\eqref{eq:c31J}.
We now consider
the first term on the r.h.s. of Eq.~\eqref{eq:c31J}. 
Taking  the $\eta_{31}\to 0$ limit in  $J^{ggg}_{S_4}$, we obtain 
\bes
C_{31} J_{S_4}^{ggg} = \qas \lp\frac{4 E_3^2}{\mu^2}\rp^{-\ep}
\frac{\Ca}{\ep^2}
\bigg[ 2\lp \eta_{12}\rp^{-\ep} K_{12} 
+\frac{\Gamma^3(1-\ep)\Gamma(1+\ep)}{\Gamma(1-2\ep)}
 \lp\eta_{31}\rp^{-\ep}\bigg],
\end{split}
\ee
where we  used 
\be
\lim_{\eta_{ij}\to 0} K_{ij} = \frac{\Gamma^3(1-\ep)\Gamma(1+\ep)}
{\Gamma(1-2\ep)}.
\ee
Since we have to apply  the $C_{31}$ operator to
$3 \la J_{S_4}^{ggg} (\I-S_3)\st \FLM(1_g,2_g,3_g)\ra_\delta$ and since the limit of $\FLM(1_g,2_g,3_g)$
is identical to what we already discussed in the NLO case, the computation
proceeds similarly to the NLO case.  Note that
since the $(E_3^2)^{-\ep}$ prefactor in $J_{S_4}$ gives an extra  factor $(1-z)^{-2\ep}$ the relevant
anomalous dimension in this case  is $\gamma^{24}_{z,g\to gg}$, c.f. 
Eq.~(\ref{eq433a}). The result of the calculation reads 
\bes
& 3\la (\I-S_3) J^{ggg}_{S_4} \st \FLM(1_g,2_g,3_g) \ra_\delta = 
\\
& \quad
+2\frac{\qas^2}{\ep^3}\lp\frac{\mu^2}{\mq^2}\rp^{2\ep}
\Ca \bigg[
\frac{2\Gamma^4(1-\ep)}{\Gamma^2(1-2\ep)}
+ 
\frac{\Gamma^4(1-\ep) \Gamma(1+\ep)}{2\Gamma(1-3\ep)}
\bigg]
\gamma_{z,g\to gg}^{24}
\la\FLM(1_g,2_g)\ra_\delta
\\
&\quad
+3\la \sum_{i=1,2} (\I-C_{3i}) \w^{3i} (\I-S_3) J^{ggg}_{S_4} 
\st \FLM(1_g,2_g,3_g) \ra_\delta.
\end{split}
\ee

We combine the  different contributions and obtain 
\bes
&12\la (\I-\SS)\st \FLM(1_g,2_g,3_g,4_g) \theta(E_3 - E_4) \ra_\delta = 
\\
&
\quad\;\;
\quad
12\la (\I-S_4)(\I-\SS)\st \FLM(1_g,2_g,3_g,4_g) \theta(E_3 - E_4) \ra_\delta 
\\
&
\quad\;\;+3 \sum_{i=1,2} \la  (\I-C_{3i}) \w^{3i} (\I-S_3) J^{ggg}_{S_4} 
\st \FLM(1_g,2_g,3_g) \ra_\delta
\\
&
\quad\;\;+2\frac{\qas^2}{\ep^3}\lp\frac{\mu^2}{\mq^2}\rp^{2\ep}
\Ca \bigg[
  \frac{2\Gamma^4(1-\ep)}{\Gamma^2(1-2\ep)}
+ 
\frac{\Gamma^4(1-\ep) \Gamma(1+\ep)}{2\Gamma(1-3\ep)}
\bigg]
\gamma_{z,g\to gg}^{24}
\la\FLM(1_g,2_g)\ra_\delta.
\end{split}
\label{eq4.35a}
\ee
In Eq.~(\ref{eq4.35a})  the third and fourth lines are free from 
 unregulated singularities 
 whereas  the second line  contains unregulated collinear singularities that need to be extracted. We explain
 how to do that in the next section.

\subsubsection{Collinear singularities: general structure}
Having regulated all the soft singularities,  we are left with only one contribution
on the right hand side of Eq.~(\ref{eq4.35a}),  
\be
12\la (\I-S_4)(\I-\SS)\st\FLM(1_g,2_g,3_g,4_g)\theta(E_3-E_4)\ra_\delta,
\ee
that still contains unregulated collinear singularities. To isolate and extract  them,
we need to introduce a partition  of unity
\be
1 = w^{31,41} + w^{32,42} + w^{31,42} + w^{32,41},
\label{eq4.37}
\ee
where $w^{3i,4j}$ are functions of the partons' emission angles. These functions  are constructed
in such a way that a product of $w^{3i,4j}$ with the matrix element squared has  non-integrable
collinear singularities if gluon $g_3$ is collinear to gluon $g_i$ or   gluon
$g_4$ is collinear to gluon $g_j$. The singularities that arise when gluons $g_3$ and $g_4$
become collinear can only occur in  the partitions $w^{31,41}$  and $w^{32,42}$.
Following Refs.~\cite{Caola:2017dug,Caola:2019nzf}, we refer to $w^{31,41}$  and $w^{32,42}$ as the triple-collinear
partitions and  $w^{31,42}$ and $w^{41,32}$ as the double-collinear partitions.   A possible choice for these
functions is given  in Appendix~\ref{app:aux}.

The double-collinear
partitions can  be dealt with in a relatively straightforward way  since the collinear singularities are clearly isolated. The
only issue that we need to address  is the construction  of a proper phase space for this contribution; we discuss how
this can be done in  Appendix~\ref{app:doubcoll}. 
For the triple-collinear partitions, we need to  order the emission angles of  gluons $g_3$ and $g_4$
and we refer to these orderings as ``sectors''  that we label as {\it a, b, c, d}, 
see Refs.~\cite{Caola:2017dug,Caola:2019nzf} for details. Explicitly, we write 
\be
\begin{split} 
1 & = \theta \left ( \eta_{41} <  \frac{\eta_{31}}{2} \right ) 
+ \theta \left ( \frac{\eta_{31}}{2} < \eta_{41}  < \eta_{31} \right ) 
+ \theta \left ( \eta_{31} <  \frac{\eta_{41}}{2} \right )
+ \theta \left ( \frac{\eta_{41}}{2} < \eta_{31}  < \eta_{41} \right ) 
\\
& = 
\theta^{(a)}  + \theta^{(b)} + \theta^{(c)}  + \theta^{(d)}.
\end{split} 
\ee
Once partitions and sectors are introduced, we can extract the
collinear limits from  the decay rates following   the procedure already discussed
for  the production case~\cite{Caola:2017dug,Caola:2019nzf}. Note, however,
that similar to the  NLO computations discussed in Section~\ref{sec:nlo},  we have  to integrate  the various splitting functions appearing in the calculation over energies
to obtain (generalized) anomalous dimensions. 

We now summarize the relevant steps for the extraction of the collinear singularities,
closely following the procedure and notation of Ref.~\cite{Caola:2017dug,Caola:2019nzf}. We introduce
the short-hand notation
\be
\G(1,2,3,4) \equiv 12(\I-\SS)(\I-S_4)\st\FLM(1_g,2_g,3_g,4_g)\theta(E_3 - E_4),
\ee
and write
\bes
\la \mathcal G(1,2,3,4)\ra_{\delta} = 
\la \G^{s_r, c_s}(1,2,3,4)\ra_\delta + 
\la \G^{s_r, c_t}(1,2,3,4)\ra_\delta +
\la \G^{s_r, c_r}(1,2,3,4)\ra_\delta.
\label{eq4.38a}
\end{split}
\ee
In Eq.~(\ref{eq4.38a}), we have introduced
\begin{itemize}
  \item 
the soft-regulated, single-collinear  contribution
\bes
\la \G^{s_r, c_s}(1,2,3,4)\ra_\delta = 
\sum_{(ij)\in\{12,21\}} \la 
\bigg[C_{3i}\df3 +  C_{4j}\df4 \bigg]\w^{3i,4j}
\G(1,2,3,4)\ra_\delta 
\\
+ \sum_{i\in\{1,2\}}
\la \bigg[\theta^{(a)} C_{4i} + \theta^{(b)} C_{43} + 
\theta^{(c)} C_{3i} + \theta^{(d)} C_{43}\bigg]
\df3\df4 \w^{3i,4i} \G(1,2,3,4)\ra_\delta;
\end{split}
\label{eq:srcs}
\ee

\item the soft-regulated triple- and double-collinear contribution, defined as
\bes
\la \G^{s_r, c_t}(1,2,3,4)\ra_\delta &= 
\sum_{i\in\{1,2\}} 
\bigg\langle \bigg[\theta^{(a)} \big[\I-C_{4i}\big]
+\theta^{(b)} \big[\I-C_{43}\big]
+\theta^{(c)} \big[\I-C_{3i}\big]
\\
&+\theta^{(d)} \big[\I-C_{43}\big]\bigg]
\df3\df4 \CC_i\w^{3i,4i}\G(1,2,3,4)\bigg\rangle_\delta
\\
&
-\sum_{(ij)\in\{12,21\}} \la C_{3i}C_{4j}\df3\df4\w^{3i,4j}\G(1,2,3,4)\ra_\delta;
\end{split}
\label{eq:srct}
\ee

\item and, finally, the soft-regulated collinear-regulated term
\bes
& \la \G^{s_r, c_r}(1,2,3,4)\ra_\delta =  \sum_{i\in\{1,2\}}
\bigg\langle
\bigg[\theta^{(a)}\big[\I-C_{4i}\big]
  +\theta^{(b)}\big[\I-C_{43}\big]
  \\
&\quad\quad +\theta^{(c)}\big[\I-C_{3i}\big]
+\theta^{(d)}\big[\I-C_{43}\big]
\bigg] \df3\df4 \big[\I-\CC_i\big] \w^{3i,4i}\G(1,2,3,4)\bigg\rangle_\delta
\\
&\quad\quad
+\sum_{(ij)\in\{12,21\}} \bigg\langle \big[\I-C_{3i}\big]
\big[\I-C_{4j}\big] \w^{3i,4j}\df3\df4 \G(1,2,3,4)\bigg\rangle_\delta.
\end{split}
\label{eq:srcr}
\ee
\end{itemize}
We remind the reader 
that the notations in Eqs.~(\ref{eq:srcs},\ref{eq:srct},\ref{eq:srcr}) are such that collinear operators act on everything
that appears
to the right of them. In particular, the notation $\la...C \df i ...\ra$ implies that a particular collinear limit should be taken in 
the phase-space element  of the parton $f_i$. More details can be found in Refs.~\cite{Caola:2017dug,Caola:2019nzf} 
where we show an explicit parametrization of the emission angles for gluons $g_3$ and $g_4$  and 
define the action
of collinear operators in terms of this  parametrization.

We discuss the terms $\la \G^{s_r,c_s}(1,2,3,4)\ra$ and $\la \G^{s_r,c_t}(1,2,3,4)\ra$ in the next two subsections.
The term $\la \G^{s_r, c_r}(1,2,3,4)\ra$ is finite and can be immediately computed in four dimensions.
This point 
is again discussed in Refs.~\cite{Caola:2017dug,Caola:2019nzf}  in the context of 
color singlet  production. Since there is no conceptual difference
between how this contribution is computed in the production and decay cases, we won't
repeat the discussion here.

\subsubsection{Soft-regulated single-collinear  contribution}
To obtain an expression for the soft-regulated single-collinear contribution $\la\G^{s_r,c_s}(1,2,3,4)\ra$
in  
Eq.~\eqref{eq:srcs},
we follow the same steps as in  the production case \cite{Caola:2017dug,Caola:2019nzf}.
After a tedious   but otherwise straightforward   computation  we obtain\footnote{We have used the 
$1\leftrightarrow2$ symmetry to obtain this formula.}
\begin{align}
&
\la\G^{s_r,c_s}(1,2,3,4)\ra_\delta = 
\frac{\qas}{\ep}
\Bigg\langle 6
\left[\lp\frac{4 E_1^2}{\mu^2}\rp^{-\ep}\gamma_{z,g\to gg}^{22}
- 2\Ca \frac{\lp4 E_4^2/\mu^2\rp^{-\ep}-\lp4 E_1^2/\mu^2\rp^{-\ep}}{2\ep}
\right]
\nonumber\\
&\quad\quad\quad
\times(\I-S_3)
\bigg[(\I-C_{32})\wt_{4||1}^{32,41} + 
\lp\frac{\eta_{41}}{2}\rp^{-\ep} (\I-C_{31})\wt^{31,41}_{4||1} 
\bigg] \st \FLM(1_g,2_g,3_g)
\Bigg\rangle_\delta
\nonumber\\
&\quad
+\frac{\qas}{\ep}
\Bigg\langle
6 \lp\frac{E_4^2}{\mu^2}\rp^{-\ep} 
(\I-S_3)
(\I-C_{31})
\big[\eta_{41}^{-\ep}\lp1-\eta_{41}\rp^{\ep}\big]\wt_{3||4}^{31,41}
\st
\nonumber\\
&\quad\quad\quad
\times\bigg[
\tilde\gamma_g(\ep) \FLM(1_g,2_g,3_g) + 
\ep \tilde\gamma_g(\ep,k_\perp) r_\mu r_\nu \FLM^{\mu\nu}(1_g,2_g,3_g)
\bigg]\Bigg\rangle_\delta
\label{eq:3.48}
\\
&\quad
+\frac{\qas^2}{\ep^2}\lp\frac{\mu^2}{\mq^2}\rp^{2\ep}
\la\FLM(1_g,2_g)\ra_\delta 
\Bigg\{
2\frac{\Gamma^2(1-\ep)}{\Gamma(1-2\ep)}
\left[\lp\gamma_{z,g\to gg}^{22}\rp^2 - 2\Ca 
\lp\frac{\gamma_{z,g\to gg}^{24}-\gamma_{z,g\to gg}^{22}}{2\ep}\rp\right]
\nonumber\\
&\quad\quad\quad
+2^{\ep}\frac{\Gamma(1-2\ep)\Gamma(1-\ep)}{\Gamma(1-3\ep)}
\bigg[\gamma_{z,g\to gg}^{22} \gamma_{z,g\to gg}^{42}
- 2\Ca 
\lp\frac{\gamma_{z,g\to gg}^{24}-\gamma_{z,g\to gg}^{42}}{2\ep}\rp
\nonumber\\
&\quad\quad\quad
+2^{\ep}\left[\gamma_{z,g\to gg}^{24}
\tilde\gamma_{g}(\ep) + \ep 
\gamma_{z,g\to gg}^{24,r}\tilde\gamma_{g,k_\perp}(\ep)\right]
\bigg]
-2\Ca 4^{\ep} \left[
\frac{\Gamma(1-2\ep)\Gamma(1-\ep)}{\Gamma(1-3\ep)} - \ep^2 \Theta_{bd}\right]
\delta_g(\ep)
\nonumber\\
&\quad\quad\quad
-4\Ca\left[\frac{\Gamma^2(1-\ep)}{\Gamma(1-2\ep)}
+
\frac{\Gamma(1-2\ep)\Gamma(1-\ep)}{2^{1-\ep}\Gamma(1-3\ep)}
-\ep^2 \Theta_{ac}\right]
\lp\frac{\gamma_{z,g\to gg}^{24}-\gamma_{z,g\to gg}^{22}}{2\ep}\rp
\Bigg\}.
\nonumber
\end{align}
 The anomalous dimensions $\gamma_{z,g\to gg}^{ij}$, that appear in 
Eq.~\eqref{eq:3.48},  are defined in Eq.~\eqref{eq433a} whereas   $\tilde\gamma_g(\ep)$,
$\tilde\gamma_{g}(\ep,k_\perp)$ and $\delta_{g}(\ep)$ can be found in 
Refs.~\cite{Caola:2017dug,Caola:2019nzf}. For completeness, we report them in Appendix~\ref{app:aux}, see Eqs.~(\ref{eq:a10},\ref{eq:a11},\ref{eq:a12}). Finally, 
$\gamma^{24,r}_{z,g\to gg}$ is defined as
\be
\gamma^{24,r}_{z,g\to gg} = \frac{29}{12}\Ca + \Ca\ep\lp\frac{371}{24} - \frac{2\pi^2}{3}\rp + 
	\Ca\ep^2\lp \frac{1559}{16} - \frac{29\pi^2}{9}- 24\zeta_3\rp + \mathcal O(\ep^3).
\ee
 Note that, as a consequence
of spin correlations,  the result in Eq.~\eqref{eq:3.48} contains a finite term  $r_\mu r_\nu \FLM^{\mu\nu}$. This term
should be understood as the corresponding matrix element squared where the polarization
vector for the gluon $g_3$ is taken to be a particular four-vector $r^\mu$.  The precise form of the vector $r$ depends
on the specific way in which the limit where gluons $g_3$ and $g_4$ become collinear is approached. 
Since we use the same parametrization of the triple-collinear phase space  as in
Refs.~\cite{Caola:2017dug,Caola:2019nzf}, the explicit
form of the vector 
$r^\mu$ can be taken from these  references.  As an example, consider the $\w^{31,41}$ partition,  
where   $p_{3}$  is written as 
\be
p_3^\mu = 
E_3 (1,\sin\theta_{31}\cos\varphi_3,\sin\theta_{31}\sin\varphi_3,\cos\theta_{31}).
\ee
Here,  $\theta_{31}$ is the relative angle between the momenta of $g_1$ and $g_3$. Upon parametrizing the collinear  limit
of $g_3$ and $g_4$ as described in Refs.~\cite{Caola:2017dug,Caola:2019nzf}, we find the following expression
for
the vector $r^\mu$ 
\be
r^\mu = (0, -\cos\theta_{31}\cos\varphi_3,-\cos\theta_{31}\sin\varphi_3,
\sin\theta_{31}).
\ee
Similar to Refs.~\cite{Caola:2017dug,Caola:2019nzf},
damping factors with tildes 
in Eq.~\eqref{eq:3.48}
indicate the damping factors computed in respective collinear limits, e.g. 
\be
\wt^{31,41}_{4||1} = \lim_{\eta_{41}\to 0} \w^{31,41}.
\ee

Finally, the  two quantities $\Theta_{ac,bd}$ in Eq.~\eqref{eq:3.48}
are the only entries  where the explicit
form of the damping factor appears  in the fully-unresolved part of the result.
They read~\cite{Caola:2017dug,Caola:2019nzf}
\bes
&\Theta_{ac} = -\la
(\I-C_{31})\left[\frac{\eta_{12}}{2\eta_{31}\eta_{32}}\right]
\wt^{31,41}_{4||1} \ln \frac{\eta_{31}}{2}\ra + \mathcal O(\ep),
\\
&\Theta_{bd} = - 2\la
(\I-C_{31})\left[\frac{\eta_{12}}{2\eta_{31}\eta_{32}}\right]
\wt^{31,41}_{3||4}\ln\frac{\eta_{31}}{1-\eta_{31}}\ra + \
\mathcal O(\ep).
\end{split}
\label{eq:bigtheta}
\ee
Taking the explicit expression for the partition functions shown Appendix~\ref{app:aux}, it is straightforward to 
obtain
\be
\Theta_{ac} = 1+\ln2+\mathcal O(\ep),~~~~~
\Theta_{bd} = 2-\frac{\pi^2}{3} + \mathcal O(\ep).
\label{eq:thetaaux}
\ee

\subsubsection{Soft-regulated triple- and double-collinear contribution}
We now discuss the triple- and double-collinear  contribution
$\la \G^{s_r,c_t}(1,2,3,4)\ra$ shown in Eq.~\eqref{eq:srct}. As indicated in the
previous section,  this term includes all  the double-unresolved
collinear contributions which arise when both gluons $g_{3,4}$ are collinear
to either gluon $g_1$ or gluon $g_2$, as well as 
single-collinear contributions where gluons $g_3$ and $g_4$ are collinear to each other.  

This contribution requires a non-trivial integration of the
triple-collinear splitting function over energies and angles of  particles that 
participate  in the splitting. The relevant computation was performed in Ref.~\cite{maxtc}.
Using the results presented there, we can write 
the final result for the soft-regulated triple- and double-collinear  contribution as
\bes
&\la\G^{s_r,c_t}(1,2,3,4)\ra_\delta = 
\\
&\quad
-\frac{\qas^2}{\ep^2}\lp\frac{\mu^2}{\mq^2}\rp^{2\ep}
\left[\lp\gamma_{z,g\to gg}^{22}\rp^{2} - 
4\Ca\lp\frac{\gamma_{z,g\to gg}^{24}-\gamma_{z,g\to gg}^{22}}{2\ep}
\rp\right]
\la\FLM(1_g,2_g)\ra_\delta
\\
&\quad
+\qas^2
\lp\frac{4 \mu^2}{\mq^2}\rp^{2\ep} 2 R^{ggg}
\la\FLM(1_g,2_g)\ra_\delta,
\label{eq4.50a}
\end{split}
\ee
where the first term on the right hand side  comes from double-collinear configurations
and the second one from the triple-collinear ones. The integral of the triple-collinear
splitting function, with soft and collinear singularities subtracted,
 is denoted by  $R^{ggg}$ in Eq.~(\ref{eq4.50a}); it  reads~\cite{maxtc} 
\bes
R^{ggg} &= 
\frac{\Ca^2}{\ep}
\lp
-\frac{1895}{216} + \frac{11\pi^2}{36} 
-\frac{11\ln2}{36} + 
\frac{2\pi^2\ln2}{3} + 
\frac{11\ln^2 2}{2}-\frac{\zeta_3}{8}
\rp
\\
&
+\Ca^2\bigg(
-\frac{335}{8}-\frac{83\pi^2}{144} + \frac{71\pi^4}{1440}
+\frac{845\ln2}{108}
+\frac{187\pi^2\ln2}{36}-\frac{169\ln^2 2}{18}
\\
&
-\frac{25\pi^2\ln^2 2}{12}
-\frac{176\ln^3 2}{9}
-\frac{\ln^4 2}{12} - 2 \Li_4\lp\frac{1}{2}\rp
+\frac{121\zeta_3}{8}
+\frac{59\ln 2 \zeta_3}{4}
\bigg).
\end{split}
\ee

\subsection{Real-virtual contribution}
We now turn to the discussion of  real-virtual contributions.  Their calculation is 
 similar to the NLO case discussed  in Section~\ref{sec:nlo}.
 As in the previous section, we illustrate the most important steps of the real-virtual
 calculation for the three-gluon final state. 
Similar to NLO, we introduce a phase-space partitioning  and write
\bes
\la\FLV(1_g,2_g,3_g)\ra_\delta = 6\la(\I-C_{31})\w^{31}(\I-S_3)\st \FLV(1_g,2_g,3_g)
\ra_\delta
\\
+ 3\la
S_3 \st \FLV(1_g,2_g,3_g)
\ra_\delta
+6
\la
C_{31}(\I-S_3)\st \FLV(1_g,2_g,3_g)
\ra_\delta.
\end{split}
\label{eq:3.56}
\ee
We note that  the $1\leftrightarrow 2$ symmetry was
used to simplify Eq.~\eqref{eq:3.56}.
The first term on the right hand side of Eq.~\eqref{eq:3.56} is fully regulated.
 The terms on the second line are  soft and  collinear subtractions, which we 
now discuss.

The starting point for the calculation of the soft subtraction contribution is the factorization property
of the one-loop amplitude~\cite{Catani:1999ss}, that leads to
\bes
&
S_3\bigg[2{\Re}\big[\mathcal M^{\rm tree}(1_g,2_g,3_g)
\mathcal M^{{\rm 1-loop},*}(1_g,2_g,3_g)\big]\bigg] = 
\left[\frac{g_s^2 e^{\ep \gamma_E}}{\Gamma(1-\ep)}\right]
 \frac{ 2\Ca (p_1\cdot p_2)}{(p_1\cdot p_3)(p_2\cdot p_3)}
\\
&\quad
\times \Bigg\{
\left[2{\Re}\big[\mathcal M^{\rm tree}(1_g,2_g)
\mathcal M^{{\rm 1-loop},*}(1_g,2_g)\big]
-\frac{\beta_0}{\ep}\lp\asontwopimu\rp |\mathcal M^{\rm tree}(1_g,2_g)|^2
\right]
\\
&\quad
-\Ca\frac{\qas}{\ep^2}\frac{\Gamma^5(1-\ep)\Gamma^3(1+\ep)}
{\Gamma^2(1-2\ep)\Gamma(1+2\ep)}
\lp
\frac{\eta_{12}}{\eta_{31}\eta_{32}}\rp^{\ep}
\lp\frac{4 E_3^2}{\mu^2}\rp^{-\ep}
|\mathcal M^{\rm tree}(1_g,2_g)|^2
\Bigg\},
\label{eq4.52a}
\end{split}
\ee
with 
\be
\beta_0 = \frac{11}{6}\Ca - \frac{2}{3}\tr\nf.
\label{eq:beta0}
\ee
The appearance of the renormalized strong coupling $g_s$ and 
of the  $\beta_0$ term  in Eq.~(\ref{eq4.52a})
is related to the fact that we work with
UV-renormalized amplitudes.
Starting from  Eq.~(\ref{eq4.52a}), we follow the discussion presented
in Section~\ref{sec:nlo} and obtain
\bes
&
3\la S_3\st\FLV(1_g,2_g,3_g)\ra_\delta = 
\\
&\quad
2\Ca \frac{\qas}{\ep^2}\lp\frac{\mu^2}{\mq^2}\rp^{\ep}
\frac{\Gamma^2(1-\ep)}{\Gamma(1-2\ep)}
\left[
\la\FLV(1_g,2_g)\ra_\delta - \frac{\beta_0}{\ep}\lp\asontwopimu\rp\la\FLM(1_g,2_g)\ra_\delta
\right]
\\
&\quad
-\frac{\Ca^2}{2} \frac{\qas^2}{\ep^4} \lp\frac{\mu^2}{\mq^2}\rp^{2\ep}
\frac{\Gamma^5(1-\ep)\Gamma^3(1+\ep)}{\Gamma(1-4\ep)\Gamma(1+2\ep)}
\la\FLM(1_g,2_g)\ra_\delta.  
\end{split}
\label{eq:4.54}
\ee

Next we consider the collinear subtraction. At one-loop, the collinear 
factorization of one-loop amplitudes  leads to  \cite{kosower-uwer}
\bes
&
C_{31} 
\bigg[2{\Re}\big[\mathcal M^{\rm tree}(1_g,2_g,3_g)
\mathcal M^{{\rm 1-loop},*}(1_g,2_g,3_g)\big]\bigg] = 
\frac{[g_s^2 e^{\ep\gamma_E}/\Gamma(1-\ep)]}{p_1\cdot p_3}\times
\\
&\quad\quad
\Bigg\{
P_{gg}\lp\frac{E_1}{E_{13}}\rp \otimes
\bigg[
2{\Re}\big[\mathcal M^{\rm tree}(13_g,2_g)
\mathcal M^{{\rm 1-loop},*}(13_g,2_g)\big]
\\
&\quad\quad\quad\quad
-\frac{\beta_0}{\ep}\lp\asontwopimu\rp |\mathcal M^{\rm tree}(13_g,2_g)|^2
\bigg]
\\
&\quad\quad
+\qas \frac{\Gamma^3(1-\ep)\Gamma(1+\ep)}{\Gamma(1-2\ep)}
{\Re}
\left[\lp-s_{13}\rp^{-\ep} P^{(1)}_{gg}
\lp\frac{E_1}{E_{13}}\rp\right]\otimes
|\mathcal M^{\rm tree}(13_g,2_g)|^2\Bigg\},
\end{split}
\label{eq:3.59}
\ee
where $s_{13} = 2 p_1 \cdot p_3 +i0$. We remind the reader that the notation 
``$13_g$'' indicates a gluon that has the same direction as the  gluon $g_1$ but whose energy $E_{13}$ is given
by  $E_{13}=E_1+E_3$. As in Sec.~\ref{sec:nlo}, the symbol  $\otimes$ in Eq.~(\ref{eq:3.59}) indicates a contraction of the one-loop spin-correlated splitting function $P_{gg}^{(1)}$
with the relevant scattering amplitudes.  The one-loop splitting function 
$P_{gg}^{(1)}$ was  computed in~Ref.~\cite{kosower-uwer}; we  report it in Appendix~\ref{app:aux} for convenience.
We note that, at variance with the production case,
the splitting function $P_{gg}^{(1)}$  is manifestly real for the decay kinematics. Following the same steps as
in  the NLO calculation described in Section~\ref{sec:nlo}, we obtain 
\bes
&
6\la C_{31}(\I-S_{3}) \st \FLV(1_g,2_g,3_g)\ra_\delta = 
\frac{\qas}{\ep}\lp\frac{\mu^2}{\mq^2}\rp^{\ep}
\frac{\Gamma^2(1-\ep)}{\Gamma(1-2\ep)}
\\
&\quad\quad
\times 2 \gamma_{z,g\to gg}^{22}
\left[\la\FLV(1_g,2_g)\ra_\delta - \frac{\beta_0}{\ep}\lp\asontwopimu\rp
\la\FLM(1_g,2_g)\ra_\delta \right]
\\
&\quad\quad
-\frac{\qas^2}{\ep}\lp\frac{\mu^2}{\mq^2}\rp^{2\ep}
\frac{\Gamma(1-2\ep)\Gamma(1-\ep)}{\Gamma(1-3\ep)}
\gamma_{z,g\to gg}^{\rm 1-loop} \la\FLM(1_g,2_g)\ra_\delta.
\end{split}
\label{eq:3.60}
\ee
In Eq.~\eqref{eq:3.60}, $\gamma_{z,g\to gg}^{\rm 1-loop}$ is the one-loop anomalous dimension,
analogous to  $\gamma_{z,g\to gg}$, obtained by  integrating
$P^{(1)}_{gg}$ in Eq.~\eqref{eq:3.59} over the energy fraction $E_1/E_{13}$. Its explicit expression is reported
in Appendix~\ref{app:aux}. 

Finally, it is also convenient to explicitly extract the  $1/\ep$-poles from the
$\FLV$ terms in Eqs.~(\ref{eq:3.56},\ref{eq:4.54},\ref{eq:3.60}). Their structure is well-known \cite{Catani:1998bh} and we  have already 
discussed it in Refs.~\cite{Caola:2017dug,Caola:2019nzf} using  our notations. For completeness, 
we report the relevant formulas below
\begin{align}
&
\la\FLV(1_g,2_g)\ra_\delta = 
-2\cos(\ep\pi)\qas\lp\frac{\mu^2}{\mq^2}\rp^{\ep}
\left[\frac{\Ca}{\ep^2} + \frac{\beta_0}{\ep}\right]
\la\FLM(1_g,2_g)\ra_\delta + 
\la\FLVfin(1_g,2_g)\ra_\delta
\nonumber\\
&
\la\FLV(1_g,2_g,3_g)\ra_\delta = \la\FLVfin(1_g,2_g,3_g)\ra_\delta
-\cos(\ep\pi)\qas 
\label{eq4.57a}
\\
&\quad
\times \left[\frac{\Ca}{\ep^2} + \frac{\beta_0}{\ep}\right]
\Bigg\langle
\left[\lp\frac{\mu^2}{s_{12}}\rp^{\ep}
+\lp\frac{\mu^2}{s_{13}}\rp^{\ep}
+\lp\frac{\mu^2}{s_{23}}\rp^{\ep}\right]
\FLM(1_g,2_g,3_g)\Bigg\rangle_\delta.
\nonumber
\end{align}
In Eq.~(\ref{eq4.57a}) the functions  $\FLVfin$ are finite in four dimensions and 
$s_{ij} = 2 p_i\cdot p_j >0$. 

\subsection{Double-virtual corrections}
The double-virtual contribution is identical to those in  the production case
described in Refs.~\cite{Caola:2017dug,Caola:2019nzf}. For convenience, we report 
the relevant formulas here. Following Ref.~\cite{Catani:1998bh}, we extract all the $\ep$-poles
from the loop amplitudes and write
\bes
&\la\FLVV(1_g,2_g)\ra_\delta = 
\lp\asontwopimu\rp^2\bigg[ \frac{\tilde I_{12}^2(\ep)}{2} - 
\frac{\beta_0}{\ep} \tilde I_{12}(\ep)
\\
&\quad\quad\quad
+\lp\frac{\beta_0}{\ep} + K\rp \frac{e^{-\ep\gamma_E}\Gamma(1-2\ep)}{\Gamma(1-\ep)} \tilde I_{12}(2\ep)
+\frac{H_g}{\ep}\bigg]\la\FLM(1_g,2_g)\ra_\delta+
\\
&\quad\quad\quad
+\lp\asontwopimu\rp \tilde I_{12}(\ep) \la \FLVfin(1_g,2_g)\ra_\delta
+\la\FLVVfin(1_g,2_g)\ra_\delta + 
\la\FLVsqfin(1_g,2_g)\ra_\delta,
\end{split}
\label{eq:3.62}
\ee
where  
\be
\tilde I_{12}(\ep) = -2\cos(\ep\pi)\left[\frac{e^{\ep\gamma_E}}{\Gamma(1-\ep)}\right]\lp\frac{\mu^2}{\mq^2}\rp^{\ep}
\left[\frac{\Ca}{\ep^2}+\frac{\beta_0}{\ep}\right],
\ee
with $\beta_0$ defined in Eq.~\eqref{eq:beta0} and
\bes
&K = \lp\frac{67}{18}-\frac{\pi^2}{6}\rp \Ca - \frac{10}{9}\tr\nf,
\\
&H_g = \Ca^2 \lp\frac{5}{12} + \frac{11\pi^2}{144} + \frac{\zeta_3}{2}\rp
-\Ca\nf\lp\frac{29}{27}+\frac{\pi^2}{72}\rp+
\frac{\Cf\nf}{2} + \frac{5\nf^2}{27}.
\end{split}
\ee
Finally, we note that $\la\FLVfin(1_g,2_g)\ra$ is defined in Eq.~\eqref{eq:245},
and $\la\FLVVfin(1_g,2_g)\ra$, $\la\FLVsqfin(1_g,2_g)\ra$ are finite 
remainders, see Appendix A in Ref.~\cite{Caola:2019nzf} for  details. 

\subsection{Final result}
\label{sechggfull}
The sum of the different contributions discussed in the previous sections  gives a
result that is finite in the $\epsilon \to 0$ limit. 
Repeating similar
calculations  for all the other partonic channels, we obtain the full NNLO QCD corrections to the decay
$H \to gg$.  We write the result as the sum of contributions with different final state multiplicities,
cf. Eq.~\eqref{eq:210}
\be
\d\Gamma^\NNLO = \d\Gamma^\NNLO_{H\to 4} + 
\d\Gamma^\NNLO_{H\to 3} + \d\Gamma^\NNLO_{H\to 2}.
\ee
The contribution of the four-parton final state reads 
\bes
& \d\Gamma^\NNLO_{H\to4} = 
\sum_{i\in\{1,2\}}
\bigg\langle
\bigg[\theta^{(a)}\big[\I-C_{4i}\big]
+\theta^{(b)}\big[\I-C_{43}\big]+\theta^{(c)}\big[\I-C_{3i}\big]
+\theta^{(d)}\big[\I-C_{43}\big]
\bigg] \times
\\
&
\quad\quad\quad\quad\quad
 \df3\df4 \big[\I-\CC_i\big] \w^{3i,4i}
 \big[\I-S_4\big]\big[\I-\SS\big] \FLM(1,2,3,4)\bigg\rangle_\delta
\\
&\quad+
\sum_{(ij)\in\{12,21\}} \bigg\langle \big[\I-C_{3i}\big]
\big[\I-C_{4j}\big] \w^{3i,4j}\df3\df4 
 \big[\I-S_4\big]\big[\I-\SS\big]
\FLM(1,2,3,4)\bigg\rangle_\delta,
\end{split}
\ee
where $\FLM(1,2,3,4)$ is defined in Eq.~(\ref{eq:419a}).  Similarly, 
the three-parton  contribution reads
\bes
\d\Gamma^{\NNLO}_{H\to 3} &= 
\bigg\langle\ONLO\mathcal J_{ggg} \big[3 \st \FLM(1_g,2_g,3_g)\big] 
+\nf \big[ 
\ONLO\mathcal J_{gqq}\; \st \FLM(1_g,2_q,3_\qb)\\
&+
\ONLO\mathcal J_{qgq}\; \st \FLM(1_q,2_g,3_\qb)+
\ONLO\mathcal J_{qqg}\; \st \FLM(1_q,2_\qb,3_g)\big]
\bigg\rangle_\delta
\\
&+\gamma_{k_\perp,g}
\la \ONLO \st r_\mu r_\nu\left[3 \FLM^{\mu\nu}(1_g,2_g,3_g) + 
\nf \FLM^{\mu\nu}(1_q,2_\qb,3_g)\right]\ra_\delta,
\end{split}
\ee
where
\be
\ONLO = (I - S_3)(I - C_{31} - C_{32})\lp\w^{31}+\w^{32}\rp, 
\ee
and 
\be
\mathcal J_{ijk} = \mathcal J_{ijk}^{(1)} + \mathcal J_{ijk}^{(2)}.
\label{eq:jfunc}
\ee
The functions   $\mathcal J^{(1,2)}$ are  defined as 
\begin{align}
\mathcal J_{ggg}^{(1)} &= \Ca \left[
\KTL_{12} + \KTL_{13} + \KTL_{23}\right] + 
\beta_0 \ln(\eta_{12}\eta_{13}\eta_{23}),
\nonumber
\\
\mathcal J_{gqq}^{(1)} & = \Ca \left[\KTL_{12}+\KTL_{13}\right] + 
(2\Cf-\Ca) \KTL_{23} 
+ (2\Cf-\Ca)\frac{3}{2} \ln(\eta_{23})
\nonumber
\\
&+ \frac{\beta_0}{2}\ln\lp\frac{E_2 E_3 \eta_{12}\eta_{13}}{E_1^2}\rp
 + \frac{3}{4}\Ca \ln\lp\frac{E_1^2\eta_{12}\eta_{13}}{E_2 E_3}\rp,
 \nonumber
\\
\mathcal J_{qgq}^{(1)} & = \Ca \left[\KTL_{12}+\KTL_{23}\right] + 
(2\Cf-\Ca) \KTL_{13} 
+ (2\Cf-\Ca)\frac{3}{2} \ln(\eta_{13})
\label{eq:4.66}
\\
&+ \frac{\beta_0}{2}\ln\lp\frac{E_1 E_3 \eta_{12}\eta_{23}}{E_2^2}\rp
 + \frac{3}{4}\Ca \ln\lp\frac{E_2^2\eta_{12}\eta_{23}}{E_1 E_3}\rp,
 \nonumber
\\
\mathcal J_{qqg}^{(1)} & = \Ca \left[\KTL_{13}+\KTL_{23}\right] + 
(2\Cf-\Ca) \KTL_{12} 
+ (2\Cf-\Ca)\frac{3}{2} \ln(\eta_{12})
\nonumber
\\
&+ \frac{\beta_0}{2}\ln\lp\frac{E_1 E_2 \eta_{13}\eta_{23}}{E_3^2}\rp
 + \frac{3}{4}\Ca \ln\lp\frac{E_3^2\eta_{13}\eta_{23}}{E_1 E_2}\rp,
 \nonumber
\end{align}
and
\bes
\mathcal J^{(2)}_{ijk} = \gamma'_i + \gamma'_j + \tilde \gamma'_k
-\wt^{31,41}_{4||1}\ln\lp\frac{\eta_{13}}{2}\rp
\lp\gamma_i + 2 C_i \ln\frac{E_3}{E_1}\rp
-\wt^{32,42}_{4||2}\ln\lp\frac{\eta_{23}}{2}\rp
\times
\\
\lp\gamma_j + 2 C_j \ln\frac{E_3}{E_2}\rp
-\left[ \wt^{31,41}_{3||4} \ln\lp\frac{\eta_{13}}{4(1-\eta_{13})}\rp
+\wt^{32,42}_{3||4} \ln\lp\frac{\eta_{23}}{4(1-\eta_{23})}\rp\right] \gamma_k,
\end{split}
\label{eq:4.67}
\ee
where $C_q=\Cf$ and $C_g = \Ca$. 
The various constants and functions used  in Eqs.~(\ref{eq:4.66},\ref{eq:4.67}) can be found in Appendix~\ref{app:aux}.

Finally, the two-parton  contribution reads
\bes
& \d\Gamma^\NNLO_{H\to 2} = 
\lp\asontwopimu\rp^2 \la\FLM(1_g,2_g)\ra_\delta \Bigg\{
\Ca^2\bigg[
\frac{65837}{324}-\frac{203\pi^2}{12}+\frac{469\pi^4}{720}
\\
 & + \ln^2 2\lp\frac{\pi^2}{6}-2\rp
-\frac{\ln^4 2}{6}-4\Li_4\lp\frac{1}{2}\rp 
+\ln2\lp 3 + \frac{11\pi^2}{9} - \frac{7\zeta_3}{2}\rp
-\frac{1859\zeta_3}{36}
\\
& +\ln\lp\frac{\mu^2}{\mq^2}\rp \lp\frac{1429}{54}-\frac{11\pi^2}{8}
-\zeta_3\rp
+\frac{203}{18}\Theta_{ac}
+\lp\frac{11\ln 2}{3}+\frac{\pi^2}{3}-\frac{131}{36}\rp\Theta_{bd}
\bigg]
\\
& +\Ca\nf \bigg[
-\frac{5701}{81} + \frac{673\pi^2}{216}
-\ln2\lp 3 + \frac{2\pi^2}{9}\rp + 2\ln^2 2
+\frac{49\zeta_3}{18}
\\
& +\ln\lp\frac{\mu^2}{\mq^2}\rp
\lp
\frac{\pi^2}{4}-\frac{15}{2}\rp
-\frac{41}{18}\Theta_{ac} + \lp\frac{23}{36}-\frac{2\ln2}{3}\rp\Theta_{bd}
\bigg]
\\
& +\Cf\nf\bigg[\frac{-27}{4} 
+ \frac{\pi^2}{6}+\frac{20\zeta_3}{3}
-\ln\lp\frac{\mu^2}{\mq^2}\rp\bigg]
+\nf^2\bigg[\frac{1889}{324}-\frac{5\pi^2}{108} + \frac{13}{27}
\ln\lp\frac{\mu^2}{\mq^2}\rp\bigg]
\Bigg\} \\
& + \lp\asontwopimu\rp \left[2\gamma'_g + \frac{2\pi^2}{3}\Ca\right]
\la \FLVfin(1_g,2_g)\ra_\delta.
\end{split}
\ee
where $\Theta_{ij}$ depends on the choice of the partition functions and are
given in Eqs.~(\ref{eq:bigtheta},\ref{eq:thetaaux}).

\section{Higgs decay to $b\bar b$}
\label{sec::nnloqq}

In this section, we consider the second type of decays,  $H \to b \bar b$.\footnote{We emphasize  that in
  this section, the Higgs boson does not couple to gluons but only to $b$-quarks. Furthermore, we assume that all quarks are massless, despite
  the $b$-quark having a non-vanishing Yukawa coupling.} The calculation
of NLO and NNLO corrections proceeds along the same lines as before but is significantly simpler. 
For this reason, we do not discuss it  and
just report the results of the calculation. Although we consider the $H\to b\bar b$ process
for definiteness, we stress that the formulas presented in this section can be applied
verbatim to other decays of color singlets to quarks, e.g.  $V\to q\bar q'$, $V = Z,W$. 

The NLO computation in this case
is simpler than for  the $H\to gg $  process discussed in Sec.~\ref{sec:nlo} because, when the
Higgs boson decays into a $b \bar b g$ final
state, singularities only arise when the gluon becomes soft and/or collinear to one of the $b$ quarks;
in other words, the $b$ quarks must be hard.  For this reason, there is no need to introduce 
the $\tilde s_{ij}$-partitioning. Repeating the NLO QCD calculation described in   Sec.~\ref{sec:nlo},
we then obtain
\bes
&\d\Gamma^\NLO_{H\to 2} = \asontwopimu \lp2\gamma'_q  + \frac{2\pi^2}{3}\Cf\rp
\la\FLM(1_b,2_{\bar b})\ra_\delta + \la\FLVfin(1_b,2_{\bar b})\ra_\delta,\\
&\d\Gamma^\NLO_{H\to3} = \la \ONLO \FLM(1_b,2_{\bar b},3_g)\ra_\delta,
\end{split}
\ee
where $\gamma'_q$ is given in Eq.~\eqref{eq:A8} and $\FLVfin(1_b,2_{\bar b})$ is a finite
virtual remainder analogous to $\FLV(1_g,2_g)$ in Eq.~\eqref{eq:245}, see Appendix A
in Ref.~\cite{Caola:2019nzf} for its explicit definition. 

At NNLO, we also do not require  any additional partitioning except perhaps 
for the $4b$ final state that arises from the
prompt decay of the Higgs boson.\footnote{To avoid confusion, we emphasize that in the previous section a 
  $4q$ final state originating  from the decay $H \to (g^*\to q \bar q) \; (g^* \to q \bar q) $  was discussed
  whereas in this section 
  we consider prompt decays to fermions. For this reason, the $4b$ final state originates from e.g. 
   $H \to (b^* \to b \bar b b ) \; \bar b$ etc.} We show in  Appendix~\ref{app:Hbbbb} that the contribution of
this subprocess to the decay rate can be written as a sum of two terms:
a term that coincides with the contribution of the  decay $H \to b \bar b  q \bar q$,  $q\ne b$, where only $b$ and $ \bar b$ can be prompt
and the $q \bar q$ pair
originates from gluon splitting,  and an interference term.  The first term can be treated without any partitioning since
the hard partons are always the two $b$-quarks.  The interference term has only a triple-collinear singularity that maps onto
the corresponding splitting function. Its proper treatment is described in
Appendix~\ref{app:Hbbbb}.

The  NNLO contribution to $H \to b \bar b$ decay is then computed following the  steps
discussed in the previous section. We 
write the NNLO contribution as a sum of  ``fixed-multiplicity'' terms $\d\Gamma^\NNLO_{Q\to i}$, $i =2,3,4$.  The four-parton
contribution  reads 
\bes
&\d\Gamma^\NNLO_{H\to4} = 
\sum_{i\in\{1,2\}}
\bigg\langle
\bigg[\theta^{(a)}\big[\I-C_{4i}\big]
+\theta^{(b)}\big[\I-C_{43}\big]+\theta^{(c)}\big[\I-C_{3i}\big]
+\theta^{(d)}\big[\I-C_{43}\big]
\bigg] \times
\\
&
\quad\quad\quad\quad\quad
 \df3\df4 \big[\I-\CC_i\big] \w^{3i,4i} 
 \big[\I-S_4\big]\big[\I-\SS\big]\mathcal F(1,2,3,4)\bigg\rangle_\delta
\\
&+\quad
\sum_{(ij)\in\{12,21\}} \bigg\langle \big[\I-C_{3i}\big]
\big[\I-C_{4j}\big] \w^{3i,4j}\df3\df4 
\big[\I-S_4\big]\big[\I-\SS\big]
\mathcal F(1,2,3,4)\bigg\rangle_\delta,
\end{split}
\ee
where now
\bes
&\mathcal F(1,2,3,4)  = \\
&\quad\quad
\left[\FLM(1_b,2_{\bar b},3_g,4_g) + 
\nf \FLM(1_b,2_{\bar b},3_q,4_\qb) + 
\FLM^{\rm int}(1,2,3,4)\right]\theta(E_3-E_4),
\end{split}
\ee
with  $\FLM^{\rm int}$ defined in Appendix~\ref{app:Hbbbb}.  The three-parton  contribution reads
\be
\d\Gamma^\NNLO_{H\to 3} = \asontwopimu
\left[
\la \ONLO \mathcal J_{qqg} \FLM(1_b,2_{\bar b},3_g)
+\gamma_{k_\perp,g}
\ONLO r_\mu r_\nu \FLM^{\mu\nu}(1_b,2_{\bar b},3_g)
\ra_\delta\right], 
\ee
where the function $\mathcal J_{qqg}$  is defined in Eq.~(\ref{eq:jfunc}) and $\gamma_{k_\perp,g}$
can be found in Appendix~\ref{app:aux}.

Finally, the two-parton  contribution reads
\bes
& \d\Gamma^{\NNLO}_{H\to 2} = 
\lp\asontwopimu\rp \lp 2\gamma'_q + \frac{2\pi^2}{3}\Cf\rp 
\la \FLVfin(1_b,2_{\bar b})\ra_\delta
\\
& +
\la \FLVVfin(1_b,2_{\bar b})\ra_\delta +\la \FLVsqfin(1_b,2_{\bar b})\ra_\delta
\\
& +\lp\asontwopimu\rp^2\la\FLM(1_b,2_{\bar b})\ra
\Bigg\{
\Cf^2\bigg[
\frac{1081}{16} -\frac{67\pi^2}{6} + \frac{2\pi^4}{3}
+9\zeta_3 + \ln\lp\frac{\mu^2}{\mq^2}\rp\times
\\
& \lp\frac{3}{4}-\pi^2+12\zeta_3\rp + 7 \Theta_{ac}
\bigg]
+\Ca\Cf\bigg[
\frac{115441}{1296}-\frac{29\pi^2}{8}-\frac{11\pi^4}{720}
+\ln^2 2\lp\frac{\pi^2}{6}-2\rp
\\
&
-\frac{\ln^4 2}{6}
+ \ln 2\lp\frac{8}{3}+\frac{11\pi^2}{9}-\frac{7\zeta_3}{2}\rp
-\frac{2135\zeta_3}{36}-4\Li_4\lp\frac{1}{2}\rp
\\
&
+\ln\lp\frac{\mu^2}{\mq^2}\rp 
\lp\frac{2329}{108}-\frac{19\pi^2}{72}-13\zeta_3\rp
+\lp\frac{11\ln2}{3}+\frac{\pi^2}{3}-\frac{131}{36}\rp \Theta_{bd}
\bigg]
\\
&
+\Cf\nf\bigg[
-\frac{9929}{648} + \frac{5\pi^2}{9} 
-\ln2\lp\frac{8}{3}+\frac{2\pi^2}{9}\rp+2\ln^2 2 + \frac{145\zeta_3}{18}
\\
&
+\ln\lp\frac{\mu^2}{\mq^2}\rp \lp-\frac{209}{54}+\frac{5\pi^2}{36}\rp
+\lp\frac{23}{36}-\frac{2\ln2}{3}\rp\Theta_{bd}
\bigg]
\Bigg\},
\end{split}
\ee
where $\Theta_{ij}$ are defined in Eqs.~(\ref{eq:bigtheta},\ref{eq:thetaaux}).

\section{Validation of results}
\label{sec::valid}

In this section, we use the analytic formulas for the fully-differential decay
rates presented above to calculate the NNLO QCD corrections to decays $H \to gg$
and $H \to b \bar b$.\footnote{For our implementation,
we take all the non-trivial amplitudes from Refs.~\cite{MCFM,Gehrmann:2005pd}.}  
We compare these results with analytic formulas extracted from 
Refs.~\cite{Schreck:2007um,Baikov:2005rw,Anastasiou:2011qx} to validate our 
calculations.\footnote{We note that similar calculations have been discussed 
earlier~\cite{Anastasiou:2011qx,DelDuca:2015zqa,Bernreuther:2018ynm,Mondini:2019gid}.}

We begin with the decay process $H \to gg$, which was discussed in Sec.~\ref{sec:nnlogg}.
We consider a Higgs boson of mass $\mq = 125$ GeV which couples to gluons through the
effective Lagrangian
\be
\mathcal{L}_{ Hgg} = -\lambda_{Hgg} H G_{\mu \nu}^{(a)} G^{\mu \nu,(a)},
\ee
where in the $\overline{\rm MS}$ scheme
\bes
\lambda_{Hgg} = -\frac{\alpha_s}{12 \pi v} \bigg\{&
1 + \left[\frac{5}{2}\Ca - \frac{3}{2}\Cf\right]
\lp\frac{\as}{2\pi}\rp
+
\bigg[
\frac{1063}{144}\Ca^2 - \frac{25}{3}\Ca \Cf + \frac{27}{8}\Cf^2
\\
&
-\frac{47}{72}\Ca \nf - \frac{5}{8}\Cf\nf
-\frac{5}{48}\Ca - \frac{\Cf}{6}
\\
&
+\ln\lp\frac{\mu^2}{m_t^2}\rp
\bigg(
\frac{7}{4}\Ca^2 - \frac{11}{4} \Ca \Cf + \Cf\nf
\bigg)
\bigg]\lp\frac{\as}{2\pi}\rp^2 + \mathcal O(\as^3)
\bigg\},
\end{split}
\ee
with $\as = \alpha_s(\mu)$ being the renormalized coupling in a theory with $5$ massless flavors
and $v$ is the Higgs vacuum expectation value, see e.g. Ref.~\cite{Grigo:2014jma}. 
For the numerical results presented below, we use $m_t=173.2$ GeV. 

For numerical checks, we split the  width for the $H \to gg$ decay into different color factors,
which allows us to check different partonic channels separately.
We write
\bes
\Gamma(H \to gg) =&\Gamma_{\rm LO}(H \to gg) + \Gamma_{\rm NLO}(H \to gg) +  \Gamma_{\rm NNLO}(H \to gg) +\mathcal O(\as^5) \\
 =&\Gamma_{\rm LO}(H\to gg) \times \biggl[1 + \left(\frac{\alpha_s}{2\pi}\right) \left(\Ca R^{(1)}_{\Ca}+\nf R^{(1)}_{\nf} \right) + \\
  &\left(\frac{\alpha_s}{2\pi}\right)^2 \left(\Ca R^{(2)}_{\Ca}+\nf R^{(2)}_{\nf}+\nf^2 R^{(2)}_{\nf^2}\right) \biggr] + \mathcal{O}(\alpha_s^5),
\end{split}
\ee
where the LO decay width that has been factored out is given by $\Gamma_{\rm LO}(H \to gg) = (\alpha_s(\mu))^2/(72 \pi^3 v^2) $.
The comparison between our results for the NLO and NNLO coefficients $R^{(1,2)}$ and those presented in Ref.~\cite{Schreck:2007um} is given in Table~\ref{tab:Hggcomp}. We present 
numerical results for a scale $\mu=2 m_H$, in order to avoid accidental cancellations between
the renormalization scale $\mu$ and the Higgs mass $\mq$ that happen for $\mu=\mq$. 
We observe agreement well below the per mille level for all coefficients.

\begin{table}
\centering
\begin{tabular}{|c|c|c|}
  \hline
  Color structure & Numerical result & Analytic result \\
  \hline
  \hline
  $R^{(1)}_{\Ca}$  & 62.749(3) & 62.749  \\
  $R^{(1)}_{\nf}$  & -3.2575(2) & -3.2575  \\
  \hline
  $R^{(2)}_{\Ca}$  & 2806.2(4) & 2806.2  \\
    $R^{(2)}_{\nf}$  & -339.63(1) & -339.63  \\
  $R^{(2)}_{\nf^2}$  & 7.4824(1) & 7.4824  \\
\hline
\hline
\end{tabular}
\caption{Comparison between numerical and analytic results for NLO and NNLO color coefficients appearing in $H \to gg$ decay. The residual Monte Carlo integration error is given in parentheses. See text for details.
\label{tab:Hggcomp}
}
\end{table}

We turn now to the decay $H \to b\bar{b}$.
Again, we consider a $125$ GeV Higgs boson and five flavors of massless quarks,
which allows us to use the results presented in  Sec.~\ref{sec::nnloqq}.
The Higgs couples to bottom quarks only, through a Yukawa interaction
\be
\mathcal{L}_{Hb\bar{b}} = - \frac{y_b}{\sqrt{2}} H b\bar{b}.
\ee
Once again, we write the result for the Higgs decay width in terms of different color structures,
factoring out the LO decay width $\Gamma_{\rm LO}(H\to b\bar{b})  = 3 y_b^2  m_H /(16\pi)$, 
\bes
\Gamma(H \to b\bar{b}) =&\Gamma_{\rm LO}(H \to b\bar{b}) + \Gamma_{\rm NLO}(H \to b\bar{b}) +  \Gamma_{\rm NNLO}(H \to b\bar{b}) + \mathcal O(\as^3) \\
=& \Gamma_{\rm LO}(H\to b\bar{b}) \times \biggl[1 + \left(\frac{\alpha_s}{2\pi}\right) \left(\Cf S^{(1)}_{\Cf}\right) + \\
  &\left(\frac{\alpha_s}{2\pi}\right)^2 \left(\Cf^2 S^{(2)}_{\Cf^2}+\Ca \Cf S^{(2)}_{\Ca \Cf}+ \tr \Cf \nf S^{(2)}_{\Cf \nf} \right) \biggr] + \mathcal{O}(\alpha_s^3).
\end{split}
\ee
The comparison between the coefficients $S^{(1,2)}$ obtained from our numerical code
and from the analytic formulas of Ref.~\cite{Anastasiou:2011qx} are displayed in Tab.~\ref{tab:Hbbcomp}.
Again, we use the scale $\mu_R = 2\mq$ for this comparison.
The agreement is consistently below the per mille level across all color structures.

\begin{table}
\centering
\begin{tabular}{|c|c|c|}
  \hline
  Color structure & Numerical result & Analytic result \\
  \hline
  \hline
  $S^{(1)}_{\Cf}$  & 12.659(2) & 12.659  \\
  \hline
  $S^{(2)}_{\Cf^2}$  & 62.59(1) & 62.60  \\
  $S^{(2)}_{\Ca \Cf}$  & 66.23(1) & 66.23  \\
  $S^{(2)}_{\Cf \nf}$  & -20.24(1) & -20.24  \\
\hline
\hline
\end{tabular}
\caption{Comparison between numerical and analytic results for NLO and NNLO color coefficients appearing in $H \to b\bar{b}$ decay.
  The residual Monte Carlo integration error is given in parentheses. See text for details.
\label{tab:Hbbcomp}
}
\end{table}

Finally, we compare  exclusive jet  rates for the $H \to b\bar{b}$ decay with those reported in Ref.~\cite{Anastasiou:2011qx}.
To do so, we use the JADE clustering algorithm with $y_{\rm cut}=0.01$ and  the distance measure defined as $y_{ij}=(p_i+p_j)^2$,
and choose the scale $\mu =\mq$.
We obtain
\bes
\Gamma_{2j}(H \to b\bar{b}) =& \Gamma_{\rm LO}(H \to b\bar{b}) \times \left[1-27.176(3) \left(\frac{\alpha_s}{2\pi}\right) -1240.8(1) \left(\frac{\alpha_s}{2\pi}\right)^2 \right] +\mathcal{O}(\alpha_s^3)\\ 
\Gamma_{3j}(H \to b\bar{b}) =& \Gamma_{\rm LO}(H \to b\bar{b}) \times \left[38.509(3) \left(\frac{\alpha_s}{2\pi}\right)  + 980.6(1) \left(\frac{\alpha_s}{2\pi}\right)^2  \right] +\mathcal{O}(\alpha_s^3) \\
\Gamma_{4j}(H \to b\bar{b}) =& \Gamma_{\rm LO}(H \to b\bar{b}) \times  376.785(8) \left( \frac{\alpha_s}{2\pi}\right)^2 + \mathcal{O}(\alpha_s^3).
\end{split}
\ee
We note that our results differ from those of Ref.~\cite{Anastasiou:2011qx} by $1\%-2\%$, which is consistent with the errors reported in that reference.
The sum of the jet rates gives the total decay rate at the scale $\mu =\mq$
\be
\Gamma(H \to b\bar{b}) =\Gamma_{\rm LO}(H \to b\bar{b}) \times \left[1+11.333(4) \left(\frac{\alpha_s}{2\pi} \right) + 116.6(2) \left(\frac{\alpha_s}{2\pi} \right)^2 \right] +\mathcal{O}(\alpha_s^3),
\ee
in excellent agreement with the analytic results at this scale
\be
\Gamma(H \to b\bar{b}) =\Gamma_{\rm LO}(H \to b\bar{b})  \times \left[1+11.333 \left(\frac{\alpha_s}{2\pi} \right) + 116.6 \left(\frac{\alpha_s}{2\pi} \right)^2 \right] +\mathcal{O}(\alpha_s^3).
\ee

Clearly, the level of numerical precision achieved for the NNLO coefficients in our calculation is
excessive since for phenomenological applications it is enough to know widths with sub-percent
accuracy. We note that to achieve this level of numerical precision within our framework, one would
typically require up to one CPU hour of computation time. 

\section{Conclusion}
\label{sec::concl}

We presented analytic formulas that describe fully-differential decays of color-singlet particles
to $q \bar q$ and $gg$ final states through NNLO QCD. The results are obtained within the nested soft-collinear
subtracted scheme that we proposed earlier in Ref.~\cite{Caola:2017dug}.  The results
 are remarkably compact and simple to implement in a numerical
code. We have validated these results by computing the NNLO QCD corrections to 
the $H \to gg$ and $H \to b \bar b$ decay
rates  and comparing them to independent numerical and analytic
computations finding per mille level agreement for observables that are known analytically. 
In addition to their phenomenological relevance
for decays of the Higgs boson and electroweak vector bosons, such as  $H \to gg, H \to b \bar b $ and $V \to q \bar q$,
these results provide an important building block for the extension of the nested soft-collinear
subtraction
scheme which will make it applicable for computations of NNLO QCD corrections to
arbitrary processes at hadron colliders.

\newpage

\section*{Acknowledgements}
The research of K.M.   was supported by the Deutsche Forschungsgemeinschaft (DFG,
German Research Foundation) under grant 396021762 - TRR 257.  The research of F.C. was partially supported
by the ERC Starting Grant 804394 \textsc{hipQCD}. F.C. would like to thank the Physics Department
of the University of Milano for hospitality during the completion of this work.

\appendix

\section{Auxiliary quantities}
\label{app:aux}
In this appendix, results for various quantities used in this paper
are summarized.  We start with discussing the partition functions. They read 
\begin{gather} 
  w^{31,41} =  \frac{\eta_{32} \eta_{42}}{d_3 d_4 }
  \left ( 1 + \frac{\eta_{31}}{d_{3421}} + \frac{\eta_{41}}{d_{3412}}  \right ),
  \\
   w^{32,42} =  \frac{\eta_{31} \eta_{41}}{d_3 d_4 } \left ( 1
  + \frac{\eta_{42}}{d_{3421}} + \frac{\eta_{32}}{d_{3412}}
  \right ), 
   \\
    w^{31,42} =  \frac{\eta_{32} \eta_{41} \eta_{43}}{d_3 d_4 d_{3412} },
   \;\;\;
   w^{32,41} =  \frac{\eta_{31} \eta_{42} \eta_{43}}{d_3 d_4 d_{3421} }, 
\end{gather}
where
\be
d_{i=3,4} = \eta_{i1} + \eta_{i2},\;\;\;\; d_{3421}  = \eta_{43} + \eta_{32} + \eta_{41},
\;\;\;\; d_{3412} = \eta_{43} + \eta_{31} + \eta_{42}, 
\ee
and 
\be
\eta_{ij}=(1 - \cos \theta_{ij})/2.
\ee
It is straightforward to check that these functions provide a partition of unity
\be
w^{31,41}  + w^{32,42} + w^{31,42} + w^{32,41} = 1.
\ee

We now present formulas for the various anomalous dimensions used in the main text. In 
our NLO discussion, we used
\be
\gamma_g = \beta_0 = \frac{11}{6}\Ca-\frac{2}{3}\tr\nf,
~~~~
\gamma'_g = \Ca\lp\frac{67}{9}-\frac{2\pi^2}{3}\rp - \frac{23}{9}\tr\nf,
\label{eq:app_gamma_g}
\ee
and
\be
\gamma_q = \frac{3}{2}\Cf,
~~~~
\gamma'_q = \Cf\lp\frac{13}{2}-\frac{2\pi^2}{3}\rp.
\label{eq:A8}
\ee
At NNLO, we also introduced
\be
\gamma_{k_\perp,g} = -\frac{\Ca}{3}+\frac{2}{3}\tr\nf,
~~~
\tilde\gamma'_g = \Ca\lp\frac{137}{18}-\frac{2\pi^2}{3}\rp - \frac{26}{9}\tr\nf,
~~~~
\tilde\gamma'_q = \gamma'_q.
\ee
Following~\cite{Caola:2017dug,Caola:2019nzf}, we defined
\begin{align}
&\tilde\gamma_g(\ep) 
= \bigg[\frac{11}{6}\Ca - \frac{2}{3}\tr\nf \bigg]
+\ep\bigg[\lp\frac{137}{18}-\frac{2\pi^2}{3}\rp \Ca - \frac{26}{9} \tr \nf
\bigg]
\label{eq:a10}
\\
&\quad
+\ep^2\bigg[
\lp\frac{823}{27} - \frac{11\pi^2}{18} -16 \zeta_3\rp\Ca 
+ \lp\frac{2\pi^2}{9} - \frac{320}{27}\rp \tr \nf\bigg] + \mathcal O(\ep^3);
\nonumber
\\
&\tilde\gamma_g(\ep,k_\perp) = 
\bigg[-\frac{\Ca}{3} + \frac{2}{3} \tr \nf\bigg]
 + \ep \bigg[-\frac{7}{9} \Ca + \frac{20}{9} \tr \nf\bigg] + \mathcal O(\ep^2);
 \label{eq:a11}
\\
&\delta_g(\ep) = 
\bigg[\lp-\frac{131}{72}+\frac{\pi^2}{6}+\frac{11}{6} \ln(2)\rp \Ca + 
\lp\frac{23}{36}-\frac{2}{3} \ln(2)\rp\tr \nf\bigg] 
\label{eq:a12}
\\
&\quad
+\ep\bigg[
\lp-\frac{1541}{216} + \frac{11\pi^2}{18} - \frac{\ln(2)}{6} + 4 \zeta_3\rp\Ca
 + 
\lp \frac{103}{54} -\frac{2\pi^2}{9} + \frac{2}{3} \ln(2)\rp \tr \nf\bigg] +
\nonumber
\\
&\quad
+\ep^2\bigg[
\lp-\frac{9607}{324} + \frac{125\pi^2}{216} + \frac{7\pi^4}{45} + \ln(2) + 
\frac{11\pi^2}{18}\ln(2) + \frac{77}{6} \zeta_3\rp \Ca 
\nonumber
\\
&\quad
+ \lp \frac{746}{81} - \frac{5\pi^2}{108} - \frac{4}{3} \ln(2) 
- \frac{2\pi^2}{9}\ln(2) - \frac{14}{3} \zeta_3\rp \tr \nf \bigg ].
\nonumber
\end{align}
In the ``gluon-only'' case, discussed in Section~\ref{sec:nnlogg},
one should set $\nf=0$ in the above formulas. 

We now discuss the one-loop gluon splitting function. 
It reads~\cite{kosower-uwer}
\bes
P^{(1),{\mu\nu}}_{gg}(z) &= 
\frac{\Ca}{\ep^2}\Bigg\{
z^{\ep}F_{21}(\ep,\ep,1+\ep,1-z)+
(1-z)^{\ep}F_{21}(\ep,\ep,1+\ep,z)
\\
&-\Gamma(1+\ep)\Gamma(1-\ep)
\left[
\lp\frac{z}{1-z}\rp^{\ep}
+
\lp\frac{1-z}{z}\rp^{\ep}
\right]-1
\Bigg\}P^{\mu\nu}_{gg}(z) 
\\
&+\frac{\nf-\Ca(1-\ep)}
{(1-\ep)(1-2\ep)(3-2\ep)}
P^{\mu\nu,{\rm new}}_{gg}(z).
\end{split}
\label{eqp1gg}
\ee
Here, $F_{21}$ is the hypergeometric function. We note that the result for the splitting
function Eq.~(\ref{eqp1gg}) is written in the conventional dimensional regularization
scheme (CDR).

The splitting functions
$P^{\mu\nu}_{gg}$, $P^{\mu\nu,{\rm new}}_{gg}$
read
\bes
&
P^{\mu\nu}_{gg} = 2\Ca\left[-g_\perp^{\mu\nu}
\lp\frac{z}{1-z}+\frac{1-z}{z}\rp+2(1-\ep)z(1-z)
\kappa^\mu_\perp\kappa^\nu_\perp\right],\\
&
P^{\mu\nu,{\rm new}}_{gg} = -2\Ca \left[1-2z(1-z)\ep\right]\kappa_\perp^{\mu}
\kappa_\perp^{\nu},
\end{split}
\ee
with $\kappa_\perp = k_\perp/\sqrt{-k_\perp^2}$.  The transversal metric tensor $g_\perp^{\mu \nu}$ and the
transversal vector
$k_\perp$ are defined relative to  the four-momentum of the collinear gluon, in the standard way~\cite{Catani:1999ss}.
 The  $d-$dimensional spin averages of the splitting functions give
\bes
&\la P^{\mu\nu}_{gg}(z)\ra = \frac{-g_{\perp,\mu\nu}}{2(1-\ep)} 
P^{\mu\nu}_{gg}(z)
 = 2\Ca\left[\frac{z}{1-z}+\frac{1-z}{z} + z(1-z)\right],
\\
&\la P^{\mu\nu,{\rm new}}_{gg}(z)\ra = \frac{-g_{\perp,\mu\nu}}{2(1-\ep)} 
P^{\mu\nu,{\rm new}}_{gg}(z) = 
-\Ca\left[\frac{1-2z(1-z)\ep}{1-\ep}\right].
\end{split}
\ee
We use these results to construct the spin-averaged splitting function
$P^{(\rm 1)}_{gg}$; we then integrate it  over $z$ to obtain the anomalous dimension $\gamma^{\rm 1-loop}_{z,g\to gg}$
following a similar procedure to the one described for the tree-level splitting function, see the
discussion leading to Eq.~\eqref{eq433a}.
We find
\bes
\gamma^{\rm 1-loop}_{z,g\to gg} &= 
\frac{11}{6}\frac{\Ca^2}{\ep^2}
+\frac{\Ca^2}{\ep}\lp\frac{134}{9}-\frac{4\pi^2}{3}\rp
+\Ca^2\lp\frac{1013}{9}-\frac{44\pi^2}{9}-60\zeta_3\rp
-\frac{\Ca\nf}{6}
\\
&+\ep\bigg[
\Ca^2\lp\frac{14635}{18}
-\frac{335\pi^2}{9}-\frac{106\pi^4}{45} - \frac{550}{3}\zeta_3\rp
-\frac{31}{18}\Ca\nf\bigg]
+\mathcal O(\ep^2).
\end{split}
\ee

Finally, the function $\KTL_{ij}$ reads 
\be
\KTL_{ij} = \Li_2(1-\eta_{ij})
+\frac{\ln^2(E_i/E_j)}{2} - \ln\lp\frac{E_i E_j}{E_3^2}\rp \ln(\eta_{ij})
+ \frac{\pi^2}{3}.
\ee

\section{Double-Collinear Phase Space}
\label{app:doubcoll}
In this appendix we describe the parametrization of the double-collinear phase space,
which turns out to be somewhat convoluted in this case.\footnote{We note
  that this issue is particular to $1 \to 2$ decays since in this case the leading
  order kinematics is overconstrained. For more complex processes, e.g.
  decays to more than two partons, $1\to N$, $N > 2$, this does not happen since
  one can always choose the angles of the two hard emittors as independent variables.}

We consider the phase space integral
\be
I=\int \df1\df2\df3\df4 (2\pi)^d \delta^{(d)}(p_H-p_1-p_2-p_3-p_4),
\label{eqdci}
\ee
with $p_H$ at rest, $p_H=(\mq,\vec{0})$. Our goal is to write the integration
measure  in Eq.~(\ref{eqdci})   in such a way that the energies $E_{3,4}$, 
the relative angle $\theta_{13}$ 
between $p_1$ and $p_3$ and the relative angle between $p_2$ and $p_4$, $\theta_{24}$,
are used as the integration variables. 
We first integrate over $\vec{p}_2$ to remove $(d-1)$ delta-functions
\be
I=\int \df1\df3\df4\frac{2\pi}{2E_2}  \delta(\mq-E_1-E_2(\vec{p}_1,\vec{p}_3,\vec{p}_4)-E_3-E_4),
\ee
with $E_2(\vec{p}_1,\vec{p}_3,\vec{p}_4)=|\vec{p}_{1}+\vec{p}_{3}+\vec{p}_{4}|$. We then
integrate over $E_1$. It is difficult to use $\cos\theta_{24}$ as an independent
variable, since $E_{2}$ is fixed by momentum conservation and thus $\cos\theta_{24}$
is a function of $E_2$ and of  $E_1$.
Instead, we parametrize the measure
in terms of $\{E_1,E_3,E_4,\vec{n}_1,\vec{n}_3,\cos\theta_{13,4}\}$ where 
$\vec n_i = \vec p_i/|\vec p_i|$ 
and $\theta_{13,4}$ 
is
the angle between the vector $\vec{p}_{13} = | \vec p_1 + \vec p_3|$ and $\vec{p}_4$,
\be
\cos\theta_{13,4}=\frac{\vec{p}_{13}}{|\vec{p}_{13}|} \cdot \vec n_4.
\ee
The integral over $E_1$ removes the remaining delta function
\be
\int \d E_1 \delta(\mq-E_1-E_2(\vec{p}_1,\vec{p}_3,\vec{p}_4)-E_3-E_4) \equiv \frac{1}{1+\frac{\partial E_2}{\partial E_1}},
\ee
where now all values of $E_1$ should be evaluated at $E_1=E_1^*$ which fulfils the $\delta$-function
constraint in the above equation. 
We obtain 
\be
I = \int \dg3 \dg4  \frac{\d \vec{\Omega}_1}{4(2\pi)^{d-2}} E_1^{-2\ep} \left[ \frac{E_1}{E_2}\frac{1}{1+\frac{\partial E_2}{\partial E_1}} \right].
\label{eq726}
\ee

We now compute $\frac{\partial E_1}{\partial E_2}$. We use
\be
E_2^2 = |\vec{p}_{13}+\vec{p}_4|^2 = |p_{13}|^2+E_4^2+2E_4 |\vec{p}_{13}| \cos\theta_{13,4},
\ee
and $|\vec{p}_{13}|^2=E_1^2+E_3^2+2E_1E_3\cos\theta_{13}$ to get
\be
\frac{\partial E_2}{\partial E_1} = \frac{E_1+E_3\cos\theta_{13}}{E_2} \left[1+\frac{E_4}{|\vec{p}_{13}|} \cos\theta_{13,4}\right].
\label{dele1bye2}
\ee
We can also rewrite the angle between vectors  $\vec p_{13}$ and $\vec p_4$ through the angle between $\vec p_2$ and $\vec p_4$.
Indeed using 
\be
\cos\theta_{13,4}=\vec{n}_{13} \cdot \vec{n}_4 = \frac{\vec{p}_{13}}{|\vec{p}_{13}|} \cdot \vec{n}_4 = -\frac{\vec{p}_{24}}{|\vec{p}_{13}|}
\cdot \vec{n}_4=-\frac{E_4+E_2 \cos\theta_{24}}{|\vec{p}_{13}|},
\label{th145th25}
\ee
in Eq.~\eqref{dele1bye2}, we find 
\be
\frac{\partial E_2}{\partial E_1} = \frac{E_1+E_3\cos\theta_{13}}{E_2} \left[1-\frac{E_4^2
    +E_2 E_4 \cos\theta_{24}}{|\vec{p}_{13}|^2} \right].
\ee
Finally, we use $|\vec{p}_{13}|^2=|\vec{p}_{24}|^2=E_2^2+E_4^2+2E_2E_4\cos\theta_{24}$ and obtain 
\be
\frac{\partial E_2}{\partial E_1} = \frac{E_1+E_3\cos\theta_{13}}{|\vec{p}_{13}|^2}  \left[E_2+ E_4 \cos\theta_{24}\right].
\ee

We now compute the energies $E_1^*$ and $E_2$ that are supposed to be used in all the formulas. 
Squaring both sides of the equation  $p_H-p_1-p_3=p_2+p_4$, we obtain 
\be
\mq^2 -2\mq(E_1+E_3)+2E_1 E_3(1-\cos\theta_{13}) = 2E_2 E_4(1-\cos\theta_{24}).
\label{d12}
\ee
We further use the energy conservation  equation $E_2 = \mq-E_1-E_3-E_4$ to find 
\be
E_1^*=\frac{\mq^2-2\mq E_3 - 2(\mq-E_3-E_4)E_4(1-\cos\theta_{24})}{2[\mq-E_3(1-\cos\theta_{13})-E_4(1-\cos\theta_{24})]}.
\ee
The energy $E_2$ is then obtained from energy conservation. 

Finally, we write the phase space for parton $f_4$ in terms of the angle $\theta_{13,4}$
\be
\df4 = \frac{\d E_4 E_4^{1-2\ep}}{2(2\pi)^{d-1}} \d \cos\theta_{13,4} (1-\cos^2\theta_{13,4})^{-\ep}\d\Omega_{13,4},
\ee
To rewrite it in terms of $\theta_{24}$, we use 
Eq.~\eqref{th145th25}
\bes
1-\cos^2\theta_{13,4} = 1-\frac{(E_4+E_2 \cos\theta_{24})^2}{|\vec{p}_{13}|^2}
=\frac{E_2^2}{|\vec{p}_{13}|^2} \left(1-\cos^2\theta_{24}\right),
\end{split}
\ee
where we applied the equality  $\vec{p}_{13}=-\vec{p}_{24}$. The Jacobian that
originates from the variable change $\cos\theta_{13,4} \to \cos\theta_{24}$ is computed  employing  Eq.~\eqref{th145th25}
one more time. The result reads 
\bes
J_{\Omega}& = \frac{\partial\cos\theta_{13,4}}{\partial \cos\theta_{24}}=-\left[ \frac{E_2}{|\vec{p}_{13}|}
  + \cos\theta_{24} \frac{\partial E_2}{\partial\cos\theta_{24}}\frac{1}{|\vec{p}_{13}|}
  - \frac{E_4+E_2\cos\theta_{24}}{|\vec{p}_{13}|^2} \frac{\partial |\vec{p}_{13}|}{\partial \cos\theta_{24}}\right].
\end{split}
\ee
To simplify it, we use
\be
\frac{\partial |\vec{p}_{13}|}{\partial \cos\theta_{24}} =  \frac{\partial |\vec{p}_{13}|}{\partial E_1}
\frac{\partial E_1}{\partial \cos\theta_{24}} = \frac{E_1+E_3 \cos\theta_{13}}{|\vec{p}_{13}|}\frac{\partial E_1}{\partial \cos\theta_{24}}.
\ee
Next we employ energy conservation
to write $\partial E_2/\partial \cos\theta_{24}=-\partial E_1/\partial \cos\theta_{24}$, and applying 
$\partial /\partial\cos\theta_{24}$ to  both sides of Eq.~\eqref{d12},  we obtain 
\be
\frac{\partial E_1}{\partial \cos\theta_{24}}=-\frac{\partial E_2}{\partial \cos\theta_{24}}=\frac{E_2 E_4}{\mq-E_3(1-\cos\theta_{13})-E_4(1-\cos\theta_{24})}.
\ee
We use this result to write the Jacobian as
\bes
J_{\Omega}=&- \frac{E_2}{|\vec{p}_{13}|}\Biggl\{1-\frac{E_4}{\mq-E_3(1-\cos\theta_{13})-E_4(1-\cos\theta_{24})}\\
&\times \left(\cos\theta_{24} +\frac{(E_4+E_2\cos\theta_{24})(E_1+E_3 \cos\theta_{13})}{|\vec{p}_{13}|^2} \right)\Biggr\}.
\end{split}
\ee
Finally, applying 
\be
\frac{E_4^2+E_2E_4\cos\theta_{24}}{|\vec{p}_{13}|^2} = 1-\frac{E_2^2+E_2E_4\cos\theta_{24}}{|\vec{p}_{13}|^2},
\ee
 we obtain the final formula for the Jacobian
\bes
J_{\Omega}&=- \frac{E_2^2}{|\vec{p}_{13}|}\frac{1}{\mq-E_3(1-\cos \theta_{13})  -E_4 ( 1 - \cos \theta_{24} ) }
\\
&\times
\Biggl[1+\frac{(E_2+E_4\cos\theta_{24})(E_1+E_3 \cos\theta_{13})}{|\vec{p}_{13}|^2}\Biggr].
\end{split}
\label{eqjack}
\ee

The  non-trivial factor present in Eq.~(\ref{eq726}) multiplied with the Jacobian in Eq.~(\ref{eqjack}) simplifies
to 
\be
\vline \frac{E_1}{E_2} J_{\Omega} \frac{1}{1+\frac{\partial E_2}{\partial E_1}} \vline
=\frac{E_1 E_2}{
|\vec p_{13}|\big[\mq-E_3 ( 1- \cos \theta_{13})  -E_4(1-\cos \theta_{24}) \big]}.
\ee
We employ this result to derive our final formula for the phase-space integral  that we use  to describe double-collinear
contributions 
\bes
I =& \int \dg3 \frac{\d E_4 E_4^{1-2\ep}}{2(2\pi)^{d-1}} \d\cos\theta_{24}
(1-\cos^2\theta_{24})^{-\ep} \left[\frac{E_1^2 E_2^2}{|\vec{p}_{13}|^2}\right]^{-\ep} 
\frac{\d\Omega_1\d\Omega_{13,4}}{4(2\pi)^{d-2}} \\
&\times
\frac{E_1 E_2}{|\vec p_{13}|\big[\mq-E_3 ( 1- \cos \theta_{13})  -E_4(1-\cos \theta_{24}) \big]}.
\end{split}
\ee
The phase space for $\dg3$ is generated using the relative angle between $\vec p_1$ and $\vec p_3$
as a variable.

\section{Prompt decays of the Higgs boson to $bb\bar b \bar b$ final  states}
\label{app:Hbbbb}
\begin{figure}[t]
\centering
\includegraphics[scale=0.12]{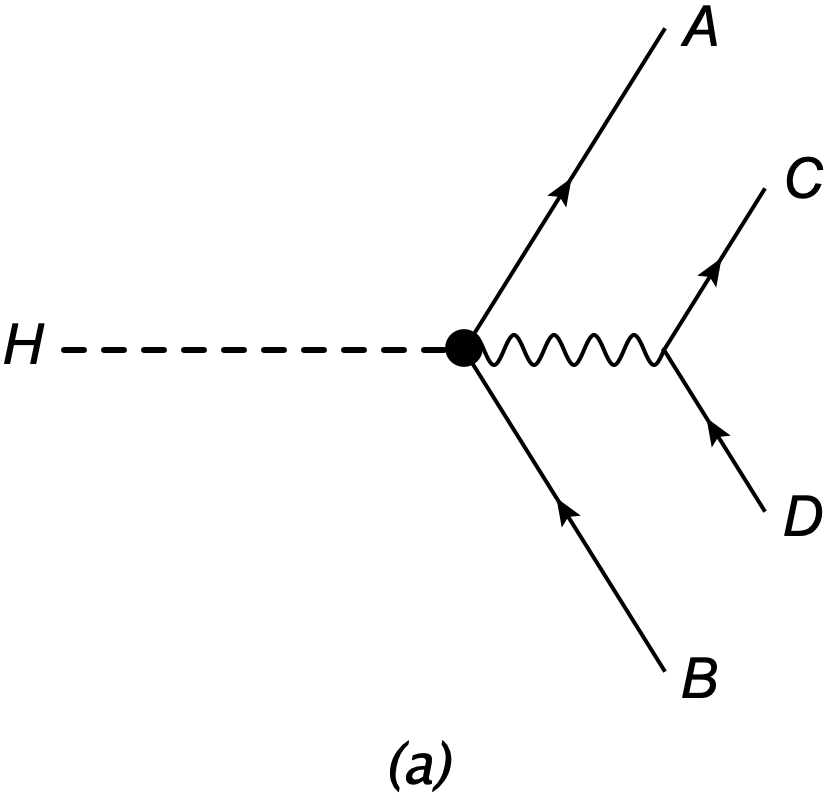}
~~
\includegraphics[scale=0.12]{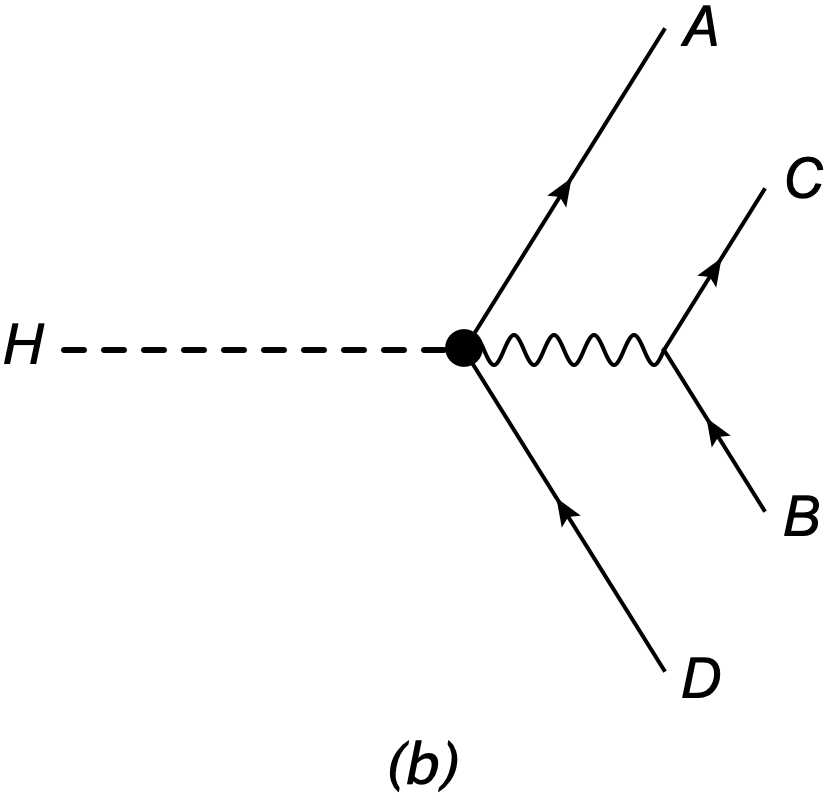}
~~
\includegraphics[scale=0.12]{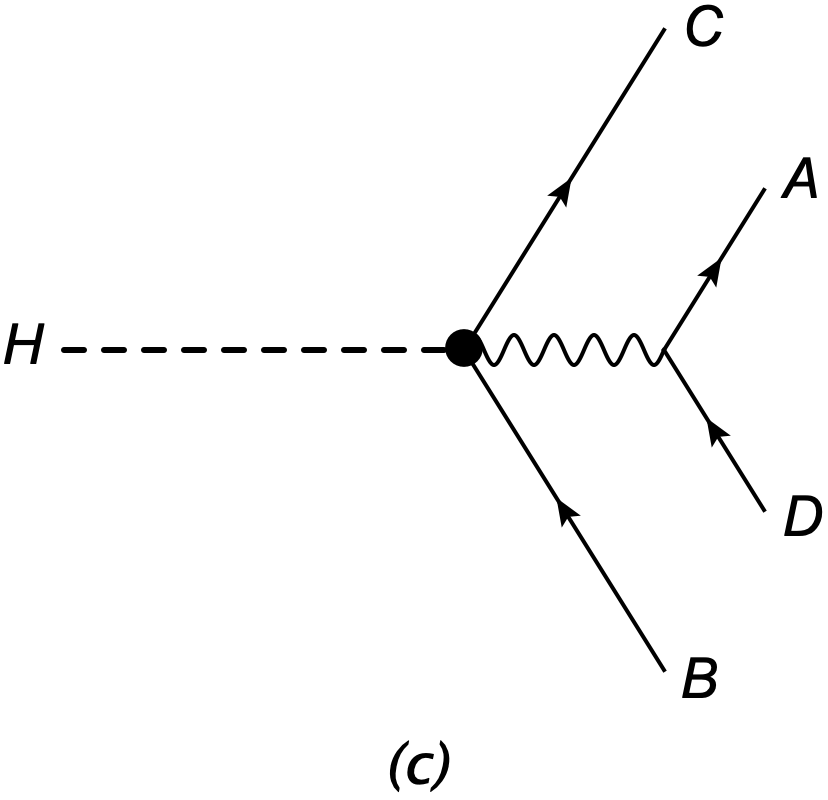}
~~
\includegraphics[scale=0.12]{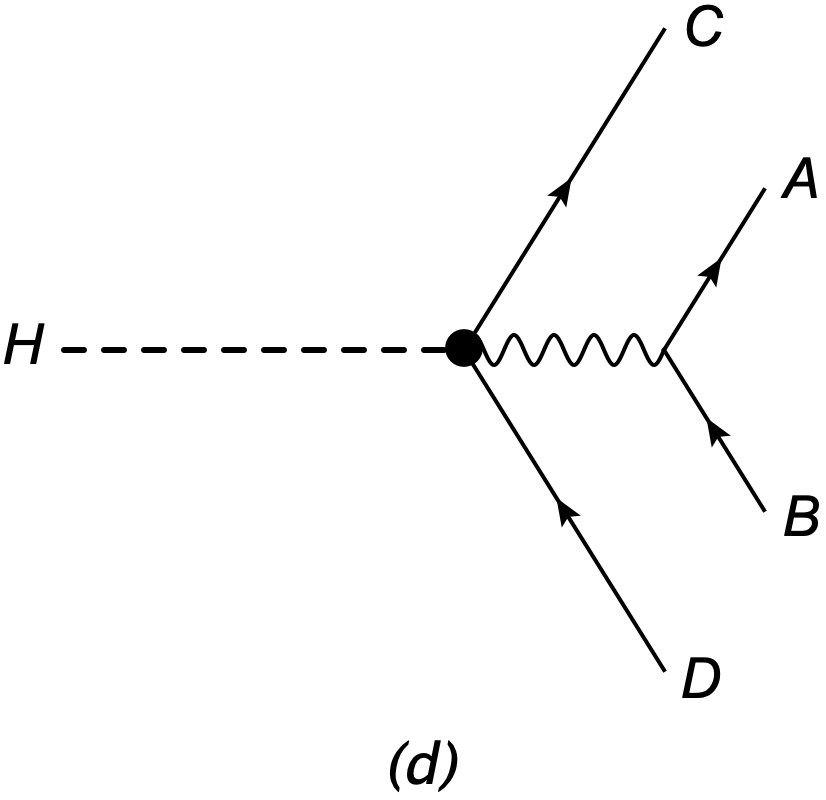}
\caption{Diagrams for $H \to b\bar{b}b\bar{b}$ decay. The gluon emission off the bubble in $Hb \bar b$ vertex
  describes  diagrams where the gluon is emitted from one of the outgoing quark lines. 
}\label{fig:bbbb}
\end{figure}

In this appendix, we consider the prompt decay of the Higgs boson\footnote{We remind the reader that in this appendix we assume
  that the Higgs boson {\it only} couples to $b$-quarks.} 
to four $b$-quarks
\be
H \to b_A + \bar b_B + b_C + \bar b_D.
\ee
There are four  subamplitudes  that contribute to this process; they are shown in  Fig.~\ref{fig:bbbb}. The difference
between these amplitudes is in the fermion lines that originate from the $Hb \bar b$ vertex and the ones that originate
from the gluon splitting, $g^* \to b \bar b$.  It is clear that $b$-quarks   from the $H \bar b b$ vertex are  hard,
in a sense that they cannot produce  infra-red singularities, 
whereas $b$-quarks
from gluon splitting can be soft. Since whether  a given $b$-quark is hard or soft 
changes from diagram to diagram,  the extraction of  singularities becomes intricate.

To overcome this problem we make use of the  symmetries of the $H\to bb\bar b\bar b$ decay.  
To this end, we write the matrix element as
the sum of four subamplitudes  shown in Fig.~\ref{fig:bbbb}
\be
\mathcal{M} = m_a + m_b + m_c + m_d
\ee
and square it.   Introducing the notation  $m_{ij}=2{\rm Re}(m_i m_j^*)$ to describe interferences of subamplitudes, 
we obtain 
\be
|\mathcal{M}|^2 =\sum_{i=a,...,d} |m_i|^2 + m_{ab} + m_{ac} + m_{ad} + m_{bc} + m_{bd} + m_{cd}.
\ee
The $H \to b\bar{b}b\bar{b}$ decay width reads 
\bes
\d \Gamma_{4b} &  \propto \frac{1}{(2!)^2} \dg A \dg B \dg C \dg D (2\pi)^4 \delta^{(4)}(p_H - p_A - p_B-p_C - p_D) \\
& \times\left\{ \sum_{i=a,...,d} 
|m_i|^2 + m_{ab} + m_{ac} + m_{ad} + m_{bc} + m_{bd} + m_{cd}\right\}.
\label{eq5.5a}
\end{split}
\ee

In the squares of amplitudes, the choice of (potentially)
hard and soft fermions  is unambiguous. We  label the hard momenta as $1$ and $2$  
and the soft momenta as $3$ and $4$. Using the symmetry of the phase space, we obtain 
\be
\begin{split} 
\d \Gamma_{4b}^{(1)} & \propto  \frac{1}{(2!)^2} \dg A \dg B \dg C \dg D
 (2\pi)^4 \delta^{(4)}(p_H - p_A - p_B-p_C - p_D) 
  \sum_{i=a,...,d} 
|m_i|^2 
\\
& = 2 \prod \limits_{i=1}^{4} \df{i}
(2\pi)^4 \delta^{(4)}(p_H - p_1 - p_2 - p_3 - p_4) 
\theta(E_3 - E_4) |m_a(1_b,2_{\bar b},3_b,4_{\bar b})|^2,
\end{split}
\label{eq5.5}
\ee
where we have included a factor of $4$ for the four diagrams  and another factor of $2$ for the energy ordering $E_3>E_4$.
It is straightforward to extract the various singularities from this contribution; in fact the result is identical
to the $q \bar q$ contribution to NNLO QCD corrections to $H \to b \bar b$ decay. 

The interference terms in Eq.~(\ref{eq5.5a}) are more involved since it is not possible 
to choose hard and soft momenta unambiguously. Before discussing this, we note that
 since helicities of massless quarks are  conserved and since $H \to b \bar b$ and $g^* \to b \bar b$
produce quarks with different (same) helicities, respectively, the interferences of diagrams $(a)$ and $(d)$ $m_{ad}$ as well
as diagrams $(b)$ and $(c)$ $m_{bc}$ vanish. We then classify
the possible collinear divergences in  the remaining interference contributions. We find the following divergences
in various interference terms:
\begin{itemize}
\item there is a triple-collinear singularity in $m_{ab}$ when $f_B,f_C$ and $f_D$ are collinear;
\item there is a triple-collinear singularity in $m_{ac}$ when $f_A,f_C$ and $f_D$ are collinear;
\item there is a triple-collinear singularity in $m_{bd}$ when $f_A,f_B$ and $f_C$ are collinear;
\item there is a triple-collinear singularity in $m_{cd}$ when $f_A,f_B$ and $f_D$ are collinear.
  \end{itemize}
Out of these four  interferences, only two are independent. The relations are 
\bes
m_{bd}(A_b,B_{\bar b},C_b,D_{\bar b}) =& m_{ac}(A_b,D_{\bar b},C_b,B_{\bar b}),\\
m_{cd}(A_b,B_{\bar b},C_b,D_{\bar b}) =& m_{ab}(C_b,B_{\bar b},A_b,D_{\bar b}).
\end{split}
\ee
We also note that
\bes
m_{ab}(A_b,B_{\bar b},C_b,D_{\bar b} ) =& m_{ab}(A_b,D_{\bar b},C_b,B_{\bar b}),\\
m_{ac}(A_b,B_{\bar b},C_b,D_{\bar b}) =& m_{ac}(C_b,B_{\bar b},A_b,D_{\bar b}).
\end{split}
\label{eq:bsymm}
\ee

\begin{figure}[t]
\centering
\includegraphics[width=0.6\textwidth]{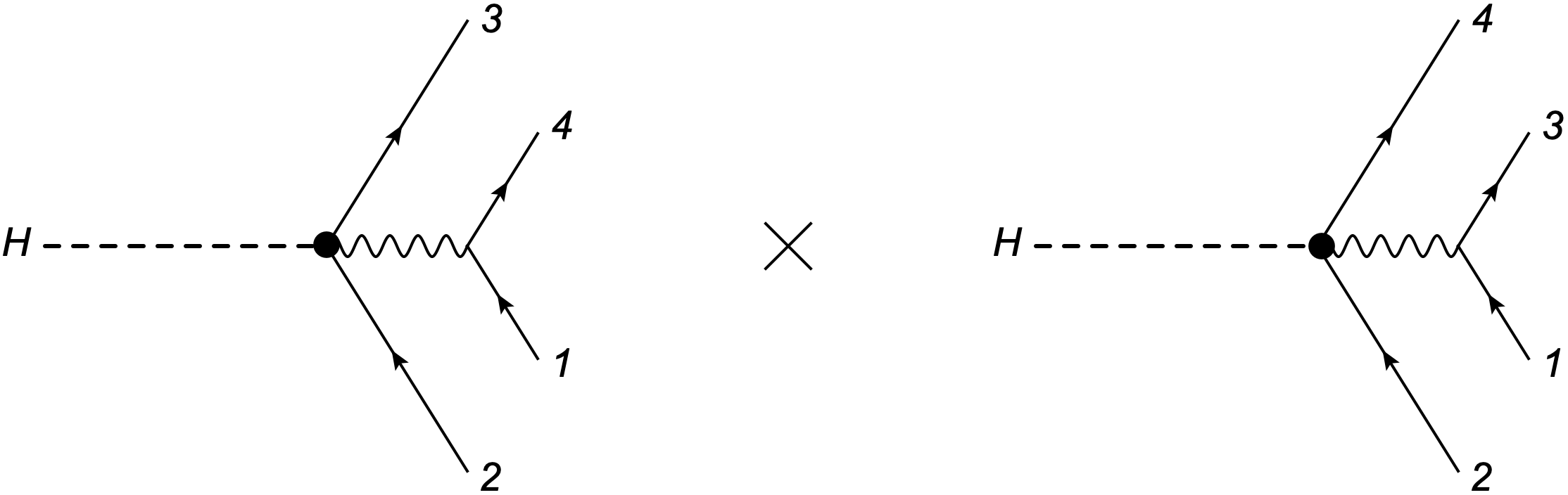}
\caption{Interference contributions for the triple collinear limit $1||3||4$.}
\label{fig:bbbbintf}
\end{figure}

Using these results, we can write
\be
\begin{split}
  \d\Gamma_{4b}  & \propto
\prod \limits_{i=1}^{4} \df{i}
(2\pi)^4 \delta^{(4)}(p_H - p_1 - p_2 - p_3 - p_4) \theta(E_3- E_4)
\\
 & \times  
\bigg\{ 2 |m_a(1_b,2_{\bar b},3_b,4_{\bar b})|^2
+ m_{ab}(2_b,3_{\bar b},1_b,4_{\bar b}) + m_{ac}(3_b,2_{\bar b},4_{b},1_{\bar b})\bigg\}.
\end{split}
\label{eq4ba}
\ee
By construction, c.f.  Fig.~\ref{fig:bbbbintf}, the interference contributions are only singular in the
limit when momenta of partons $f_1,f_3$ and $f_4$ become collinear,
whereas the non-interference term has multiple singularities,
including the double-soft one that occurs when
$f_3$ and $f_4$ become soft.   We denote  the interference term as
\be
\langle\FLM^{\rm int}(1,2,3,4)\rangle_{\delta} =
\langle \theta(E_3- E_4) \{ m_{ab}(2_b,3_{\bar b},1_b,4_{\bar b}) + m_{ac}(3_b,2_{\bar b},4_{b},1_{\bar b}) \} \rangle_\delta,  
\ee
and  use this notation in the main text when we discuss the computation of
NNLO QCD contribution to Higgs decay to two quarks in
Section~\ref{sec::nnloqq}. The non-interference term in Eq.~(\ref{eq4ba}) is accounted for as part of the
$n_f$-dependent contributions in the NNLO QCD computation.

\end{document}